\documentclass[iop,numberedappendix]{emulateapj}
\usepackage{hyperref}

\defcitealias{Zhong2015a}{Z15}

\hypersetup{
pdftitle={Selecting M-giants with infra-red photometry},
pdfauthor={Jing Li},
colorlinks=true,
linkcolor=blue,
citecolor=blue,
urlcolor=blue
}

\begin{document}

\title{Selecting M-giants with infra-red photometry: Distances,
metallicities and the Sagittarius stream}
\shorttitle{Selecting M-giants with infra-red photometry}
\shortauthors{Li et al.}

\author{
Jing Li\altaffilmark{1,2},
Martin C. Smith\altaffilmark{1},
Jing Zhong\altaffilmark{1},
Jinliang Hou\altaffilmark{1},
Jeffrey L. Carlin\altaffilmark{3},
Heidi Jo Newberg\altaffilmark{3},
Chao Liu\altaffilmark{4},
Li Chen\altaffilmark{1},
Linlin Li\altaffilmark{1},
Zhengyi Shao\altaffilmark{1},
Emma Small\altaffilmark{1},
Hao Tian\altaffilmark{5}
}

\altaffiltext{1}{
Key Laboratory of Galaxies and Cosmology, Shanghai Astronomical
Observatory, Chinese Academy of Sciences, 80 Nandan Road, Shanghai,
200030, China; lijing@shao.ac.cn, msmith@shao.ac.cn}
\altaffiltext{2}{University of Chinese Academy of Sciences, 19A Yuquan Road, Beijing 100049, China}
\altaffiltext{3}{Dept. of Physics, Applied Physics and Astronomy, Rensselaer Polytechnic Institute Troy, NY 12180}
\altaffiltext{4}{Key Laboratory of Optical Astronomy, National Astronomical Observatories,Chinese Academy of Sciences, Datun Road 20A, Beijing 100012, China}
\altaffiltext{5}{Kapteyn Astronomical Institute, University of Groningen, P.O. Box 800, 9700 AV Groningen, The Netherlands}

\begin{abstract}
Using a spectroscopically confirmed sample of M-giants, M-dwarfs and
quasars from the LAMOST survey, we assess how well WISE $\&$ 2MASS
color-cuts can be used to select M-giant stars. The WISE bands are
very efficient at separating M-giants from M-dwarfs and we present a
simple classification that can produce a clean and relatively complete
sample of M-giants. We derive a new photometric relation to estimate
the metallicity for M-giants, calibrated using data from the APOGEE
survey. We find a strong correlation between the $(W1-W2)$ color and
$\rm [M/H]$, where almost all of the scatter is due to photometric
uncertainties. We show that previous photometric distance relations,
which are mostly based on stellar models, may be biased and devise a
new empirical distance relation, investigating trends with
metallicity and star formation history. Given these relations, we
investigate the properties of M-giants in the Sagittarius stream. The
offset in the orbital plane between the leading and trailing tails is
reproduced and, by identifying distant M-giants in the direction of
the Galactic anti-center, we confirm that the previously detected
debris in the outer halo is the apocenter of the trailing tail.
We also find tentative evidence supporting an existing
overdensity near the leading tail in the Northern Galactic hemisphere,
possibly an extension to the trailing tail (so-called Branch C).
We have measured the metallicity
distribution along the stream, finding a clear metallicity offset
between the leading and trailing tails, in agreement with models for
the stream formation. We include an online table of M-giants to
facilitate further studies.
\end{abstract}
\keywords{
stars: late-type --
stars: distances --
stars: abundances --
Galaxy: structure --
Galaxy: halo --
galaxies: individual (Sagittarius)
}

\section{Introduction}
\label{sec:intro}

Standard $\Lambda CDM$ cosmology predicts that the stellar halo is
formed by merger of smaller satellite systems. This process would
leave its imprint in a diffuse stellar halo on a large scale.
The density, luminosity, velocity, and metallicity of the stellar
halo can thus provide direct constraints on the evolution history
(e.g. \citealt{nyetal02}; \citealt{rochapinto2004};
\citealt{Belokurov2006}; \citealt{heidi2009}; \citealt{Smith2009};
\citealt{Schlaufman2012}; \citealt{Drake2013}).
Therefore detecting structures in the stellar halo of the Milky Way is
particularly important for understanding the evolution of our Galaxy.

M-giant stars, with temperatures in the range 2,400 to 3,700 K, are
very bright; at the tip of the red giant branch, their luminosities
are $\sim10^3$~L$_{\odot}$, making them excellent tracers of
substructures at large distances in the outer Galactic halo
\citep[][etc.]{Majewski2003,Sharma2010,Bochanski2014,Sheffield2014}. However,
only relatively metal-rich stellar populations in the red giant branch
(RGB) can extend to cool enough temperatures to produce M-giants,
making M-giants a rare spectral type in the mostly metal-poor Galactic
halo.

As M-giants are useful for tracing metal-rich structures, they are
ideal for tracing the Sagittarius (Sgr) stream. Since this stream is
(relatively) metal-rich it contains a significant number of
M-giants.
Currently, M-giants are mainly selected from multi-band photometric
surveys as spectral identification is only available for small
samples. From a Near Infra-Red (NIR) color-color diagram of stars from
2MASS photometry catalog, \citet{Majewski2003}, for the first time,
selected thousands of M-giant candidates which were used to map the
all-sky view of the Sgr dwarf galaxy. Recently,
\citet{Bochanski2014} used a combined NIR and optical color
selection to assemble a catalog of 374 faint M-giant candidates from
UKIDSS and SDSS. These candidates were selected by position in the
$(J-H)_{0}$ vs $(J-K)_{0}$ color-color diagram for point sources in
UKIDSS. This NIR color-color selection is similar to that of
\citet{Majewski2003}, but slightly shifted reward to reduce
contamination from bluer nearby M-dwarfs.
They used a NIR-optical $(g-i)_{0}$ vs $(i-K)_{0}$ color-color
selection to identify quasar (QSO) contamination, removing nearly half
of their M-giant candidates.
In order to further confirm those M-giants, they have begun high
resolution spectroscopic follow-up observations. From their sixteen
M-giant candidates only 3 are confirmed to be genuine M-giants, which
implies that the efficiency of selecting true M-giants is only about
$\sim18\%$, although this is partially a consequence of operating at
faint magnitudes where the fraction of M-giants to M-dwarfs drops
dramatically.
During the preparation of this paper, \citet{Koposov2014} showed that
mid-IR photometry is more effective for classification and used this
technique to investigate Sgr debris towards the Galactic anti-center.

It is clear that M-giant samples need to be enlarged, not only
with respect to the number of candidates, but also in the sky
coverage. The selection efficiency also needs to be improved
substantially. An important way to test and improve photometric
selection techniques is through large spectroscopic surveys containing
confirmed M-giant stars. One such survey is the LAMOST Experiment for
Galactic Understanding and Exploration (LEGUE). The LAMOST telescope
(Large sky Area Multi-Object fiber Spectroscopic Telescope, also known
as the Guo Shou Jing Telescope) has been in full survey mode since
2012 \citep{Luo2015}.
From the huge catalog of LAMOST spectra \citeauthor{Zhong2015a}
(\citeyear{Zhong2015a}, hereafter \citetalias{Zhong2015a}; see
also \citealt{Zhong2015b}) constructed M-giant
templates and identified around 9,000 M-giant stars, which is
currently the largest spectroscopic catalog of M-giants. 
This large spectroscopically-selected catalog provides a mechanism
with which we can explore more efficient multi-band photometric
methods for separating M-giants from the much more numerous M-dwarfs
in the Milky Way.

The paper is arranged as follows. Section \ref{sec:LAMOST} briefly
describes the LAMOST data we used in this work and the spectroscopic
M-giant/M-dwarf identification. In Section \ref{sec:photometry} we
recap previous methods for selecting M-giants and introduce our new
selection criteria based on IR photometry. In Section
\ref{sec:properties} we analyze the properties these M-giants,
constructing photometric relations for estimating metallicities
(\ref{sec:metallicity}) and distances (\ref{sec:distance}). In Section
\ref{sec:sgr} we use our selection criteria to identify M-giants from
the all-sky ALLWISE database in order to study the Sgr stream. We give
our conclusions in Section \ref{sec:conclusions}.

\section{LAMOST data and spectral classification}
\label{sec:LAMOST}

In order to study the properties of M-giants, we first need to
identify a large sample of such stars. We do this using data from the 
LAMOST spectroscopic telescope \citep{Zhao2012}. LAMOST is a
large-aperture (4-5m) spectroscopic survey telescope, operated by
National Astronomical Observatories, Chinese Academy of Sciences. It
has a total of 4,000 fibers across a 20 sq. deg. field-of-view. Since
2012 the telescope has been carrying out an extensive survey of stars
in the Milky Way \citep{Deng2012}, obtaining spectra for over 1
million stars each year at a resolution of ${\rm R} \approx
1,800$. In general, the input catalog uses a selection method which
is either random or preferentially selects less populous regions of
color-magnitude space \citep{Luo2015}. DR1 was released to the
international community in 2015 \citep{Luo2015}.

The survey has been shown to be extremely useful for
obtaining large samples of red giant stars \citep[notably the sample
of K-giants from][]{Liu2014}, but unfortunately the LAMOST pipeline
has difficulties with cool M-type stars, in particular in the 2D
spectral analysis and flux calibration \citep{Luo2015}. This leads to
problems classifying late-type stars such as M-giants due to the
numerous molecular bands which affect the morphology of the spectra.

To circumvent these problems a separate analysis of the M-type stars
needs to be undertaken. For example, \citet{Yi2014} applied a
modification of the Hammer technique \citep{Covey2007} to a set of
spectra from the LAMOST pilot survey, obtaining a sample of 67,000
M-dwarfs.

Another analysis has been carried out by \citetalias{Zhong2015a} and it
is this sample which we will use as a foundation for our study.
\citetalias{Zhong2015a} selected a large sample of M-giants and
M-dwarfs from LAMOST Data Release 1 (DR1) using a template fitting
method to select M-type stars. They find over 100,000 spectra
that show the characteristic molecular titanium oxide (TiO), vanadium
oxide (VO) and calcium hydride (CaH) features typical of M-type
stars. They then take this sample and calculate the TiO5, CaH2, and
CaH3 spectral indices. The TiO and CaH spectral indices
were defined by \citet{Reid1995} and \citet{Lepine2007}, and the
distribution of the spectral indices is a good indicator to separate
M-dwarf stars with different metallicity
\citep{Gizis1997,Lepine2003,Lepine2007,Lepine2013,Mann2012}.
Figure 1 in \citetalias{Zhong2015a} shows the distribution of
identified M-type stars in the TiO5 against CaH2+CaH3 plot. The
M-giants generally have weaker CaH molecular bands for a given range
of TiO5 values, which is also consistent with the giant/dwarf
discrimination of \citet{Mann2012}. We note that the spectral indices
distribution of M-type stars can be used to separate giants from
dwarfs with little contamination, especially for late-type M-giants.

The analysis of \citetalias{Zhong2015a} resulted in two catalogs, one
containing 8,639 M-giant candidates and one with around 100,000 M-dwarf
candidates. It is these samples which we use in the following
study. We augment these with a further catalog of 4,000 spectroscopically
confirmed QSOs from LAMOST DR1, in order to assess QSO contamination.

\section{Photometric selection of M-giants}
\label{sec:photometry}

\subsection{WISE and 2MASS photometric data}
\label{sec:WISE}

With these relatively pure spectroscopically identified M-giants,
M-dwarfs and QSO candidates selected from LAMOST DR1,
we now attempt to find an effective selection criteria for
M-giants using near- and mid-IR photometry.

The Wide-field Infrared Survey Explorer (WISE) is a medium class space
Explorer mission funded by NASA and launched in December 2009. The
WISE project mapped the whole sky in four bands ($W1$, $W2$, $W3$ and
$W4$), centered at wavelengths of 3.4, 4.6, 12 and 22 $\mu$m
\citep{Wright2010}.

We cross-matched our LAMOST sample of M giants, M dwarfs and QSOs
to the ALLWISE Source Catalog in the NASA/IPAC Infrared Science
Archive, using a search radius of $3\arcsec$.
More than 99\% of the matched objects had LAMOST-WISE source
separations less than $1\arcsec$. We also applied the following cuts
to ensure high quality data: $cc_{\rm flags}=0000$; $ext_{\rm flg}=0$;
${\it W1}_{\rm sigmpro}$ \& ${\it W2}_{\rm sigmpro} <0.05$, which
means that the profile-fit photometric uncertainty is less than
$0.05$ mag; and finally three cuts to ensure that we avoid saturated
photometry, ${\it W1}_{\rm sat}$ \& ${\it W2}_{\rm sat}< 0.05$ (which
means that the saturated pixel fraction is less than 0.05), and
${\it W1}>8.1$ mag (corresponding to the ${\it W1}$ saturation magnitude).
After applying these quality cuts, we are left with 4,136 M-giants,
63,979 M-dwarfs, and 644 QSOs with both LAMOST and WISE $\&$ 2MASS
data. The mean photometric errors for our sample are:
$\delta J = 0.023$ mag, $\delta K = 0.024$ mag, $\delta W1 = 0.023$
mag, and $\delta W2 = 0.021$ mag.

Throughout we adopt the $E(B-V)$ maps of \citet{Schlegel1998}, in
combination with $A_r/E(B-V) = 2.285$ from \citet{Schlafly2011} and
$A_\lambda/A(r)$ from \citet{Davenport2014}. This final conversion has
a weak dependence on latitude (see Section 4.2 of their
paper). In Appendix \ref{sec:extinction} we describe our
implementation of this latitude dependence. For illustrative purposes,
in the following figures we include an arrow denoting the reddening vector.
For this we choose $b=30$ deg, which corresponds to the value for high
galactic latitudes. At high galactic latitudes the orientation of the
reddening vector is independent of b; at smaller latitudes the
dependence means that in the $E(J-K)$ vs $E(W1-W2)$ plane the 
reddening vector becomes closer to vertical as $\left|b\right|$
approaches 0.
The length of the reddening vector in these figures corresponds to an
extinction of $A_J = 0.5$ mag; this is around an order-of-magnitude
larger than typical values at high latitudes, but underestimates the
extinction at low latitudes (i.e. $\left|b\right| \la 10$ deg) and
towards the central parts of the Galaxy.

\subsection{Previous selection methods}

In the past decades many studies have used M-giants to trace
substructures in the Milky way halo
\citep[e.g.][]{Majewski2003, Yanny2009, Rocha2003, Belokurov2014}.
One issue for these studies is the problem of contamination from
M-dwarfs, especially for samples derived from photometry. In this
subsection we discuss previous methods which have been used to select
M-giants.

The simplest and most robust way to identify M-giants is to use
features in their spectra. In the optical band the main differences
between giants and dwarfs can be seen in the morphology near the TiO
band heads and also the lack of strong Na I absorption at ~8200
\AA. In the $K$ band, giants possess significantly stronger CO bands
than their dwarf counterparts \citep{Bochanski2014}.
Another method, also based on spectroscopy, is to use the stellar
parameters (such as effective temperature and surface gravity) to
separate M-giants. An example of this approach is shown in figure 4 and
Equation (2) of \citet{Belokurov2014}, although in this example the
temperature range they have selected is more-commonly associated to
K-type stars. This was a deliberate decision due to issues with
the reliability of the spectroscopic temperature estimation for M-type
stars (V. Belokurov, private communication).
As we will show later in Section \ref{sec:2MASS_WISE}, 2MASS colors
can be used to differentiate spectral class. The colors of stars
from the \citeauthor{Belokurov2014} selection ($0.7\la(J-K)_0\la0.9$
mag) are consistent with early-type M-giants, although some may be
late-type K-giants.
Clearly the efficacy of this method relies on the accuracy of the
measurements of temperature and gravity.

\begin{figure}
\plotone{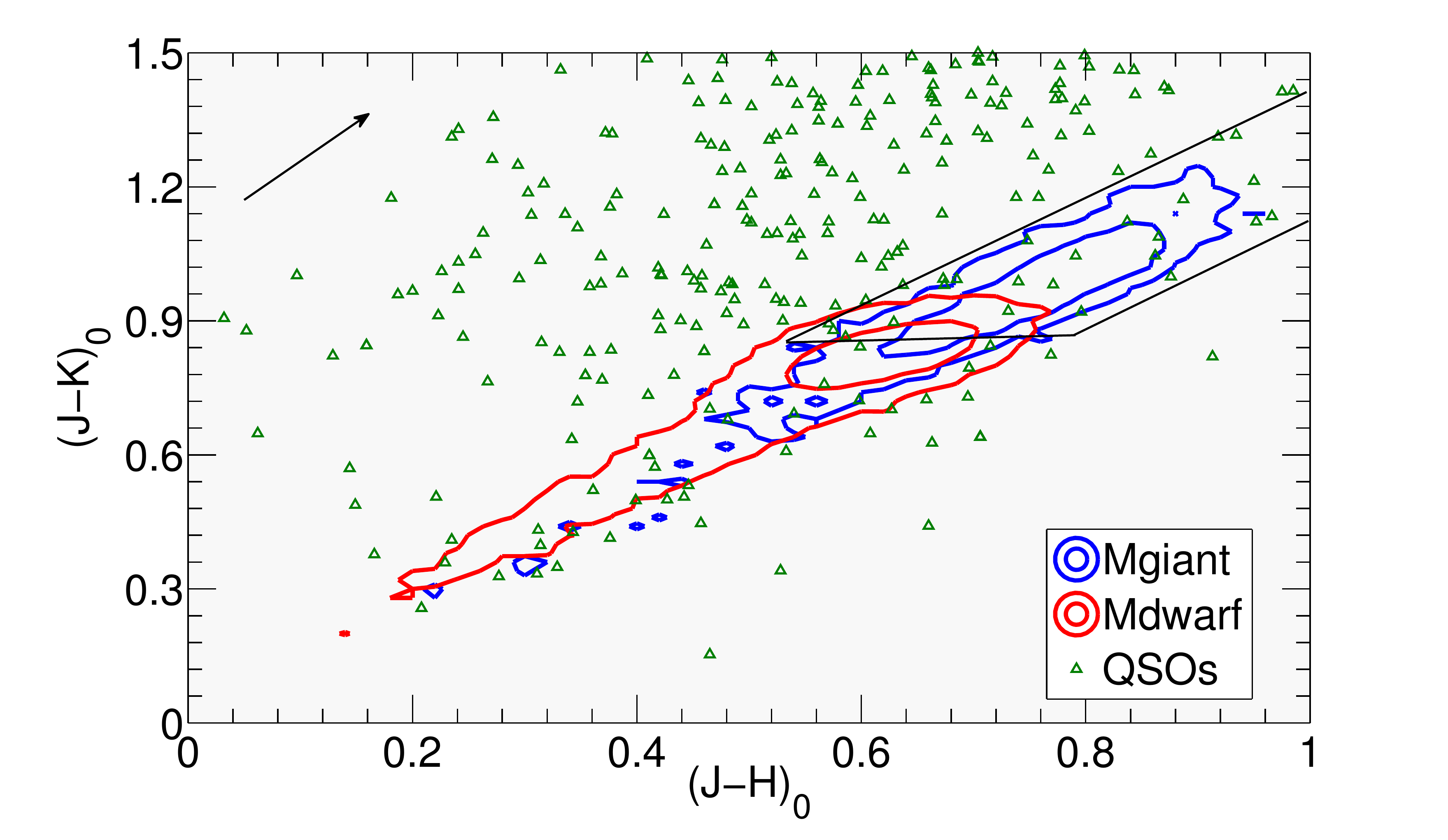}
\caption{2MASS $(J-K)_0$ vs $(J-H)_0$ color-color distribution for
spectroscopically classified M-giants (blue contours), M-dwarfs (red
contours) and QSOs (green triangles). The contours correspond to 1-
and 2-$\sigma$, i.e. contain 68.3\% and 95.4\% of the data), the.
The black arrow shows an example reddening vector (see Section
\ref{sec:WISE} for further details) and the black box shows
\citet{Majewski2003}'s selection criteria.}
\label{2mass_CC_distribution}
\end{figure}

The above two methods both rely on spectroscopic data, but to
efficiently map wide areas (and large volumes) it is often better to
use photometric surveys. This can be done using optical data
\citep[e.g.][]{Yanny2009}, although better results have been achieved
using NIR data, such as 2MASS \citep[e.g.]{Majewski2003, Sharma2010, Bochanski2014}. 
Figure~\ref{2mass_CC_distribution} shows the 2MASS color-color diagram
for our sample of M-giants, M-dwarfs and QSOs identified from LAMOST
(see Section \ref{sec:LAMOST}).
The blue contours show the distribution of M-giants (corresponding to
1- and 2-$\sigma$, i.e. containing 68.3\% and 95.4\% of the data), the
red contours M-dwarfs, and the green triangles QSOs. The black arrow
shows an example reddening vector (see Section \ref{sec:WISE} for
further details).

The stellar locus is clearly defined in this figure, but there is no
obvious division between M-giants and M-dwarfs. Although the QSOs are
offset from the stars, there is also significant overlap.
In \citet{Majewski2003}, they used $(J-K)_0>0.85$ to
remove M-dwarf contamination and the black box in
Figure~\ref{2mass_CC_distribution} shows their final selection
region. Although these color cuts remove many of the M-dwarf and QSO
contaminants, there are clearly still a significant number of M-dwarfs
present in this region. In \citet{Bochanski2014}, they used a similar
selection region, applying a stricter color cut of $(J-K)_0>1.02$ to
remove more M-dwarf contamination. This indeed removes more of the
M-dwarfs, but at a cost of removing many genuine M-giants with
$(J-K)_0<1.02$. After applying the \citet{Bochanski2014} color cuts to
our spectroscopically classified sample, we find that these
2MASS-selected M-giant candidates still have 13.5\% dwarf
contamination and many M-giants have been sacrificed (the M-giant
completeness is only 48.7\%). Although these percentages will depend
on the magnitude range under consideration and the underlying
LAMOST selection-function, they give an indication of the problems
faced by these techniques. In short, by using NIR data it is not
possible to entirely remove M-dwarf contamination and simultaneously
ensure reasonable levels of M-giant completeness.

\subsection{2MASS $\&$ WISE selection criteria }
\label{sec:2MASS_WISE}

\begin{figure}
\plotone{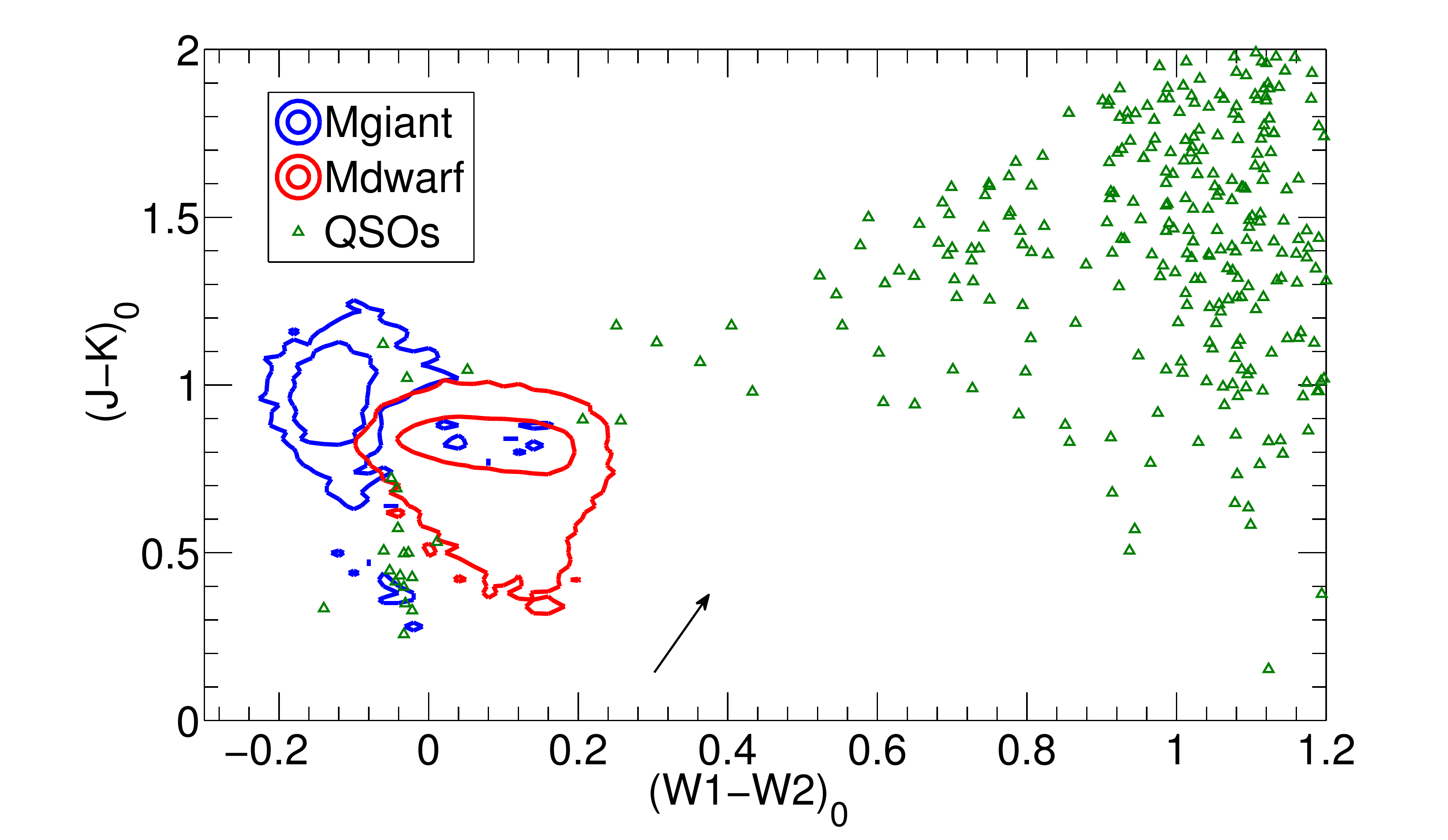}
\caption{2MASS and WISE $(J-K)_0$ vs $(W1-W2)_0$ color-color
  distribution for spectroscopically classified M-giants (blue
  contours), M-dwarfs (red contours) and QSOs (green triangles). The
  contours corresponding to 1- and 2-$\sigma$. 
  The black arrow shows an example reddening vector (see Section
  \ref{sec:WISE} for further details).
  There are a small number of M-giant located in the
  M-dwarf region, but these are most likely mis-classified due to
  having low S/N spectra (see Section \ref{sec:2MASS_WISE}.}
\label{wise_CC_distribution}
\end{figure}

A better way to identify M-giants is to use mid-IR data, such as that
available from the WISE mission. A combined 2MASS and WISE color-color
diagram for  M-giants, M-dwarfs and QSOs is shown in
Figure~\ref{wise_CC_distribution}, where the data used here are same
as in Figure~\ref{2mass_CC_distribution}.

The separation of QSOs from stars is substantially improved, 
with the former having much redder $({\it W1}-{\it W2})_{0}$ color.
As before the separation is not perfect and
a few QSOs appear to lie amid the stellar locus, particularly around
$(J-K)_0 \approx 0.5$. Although most of these are low S/N spectra
(almost all have ${\rm S/N < 10}$) and hence are probably
mis-classified, around 20\% or more appear to be binary stars composed
of an M-dwarf around a hotter companion.

From this figure one can also see that the $(J-K)_0$ vs
$({\it W1}-{\it W2})_{0}$ color-color plane enables a relatively clean
separation of M-giants from M-dwarfs, much more efficiently than NIR data
alone. There appears to be a small number of M-giants located
in M-dwarf region, but these are actually misidentified M-giants. For low
S/N spectra the dwarf/giant spectroscopic classification breaks down
and it becomes increasingly difficult to separate the two
populations. If we remove stars with low S/N spectra (i.e. ${\rm S/N <
5}$), then these spurious objects almost all disappear.

\begin{figure}
\includegraphics[width=\columnwidth]{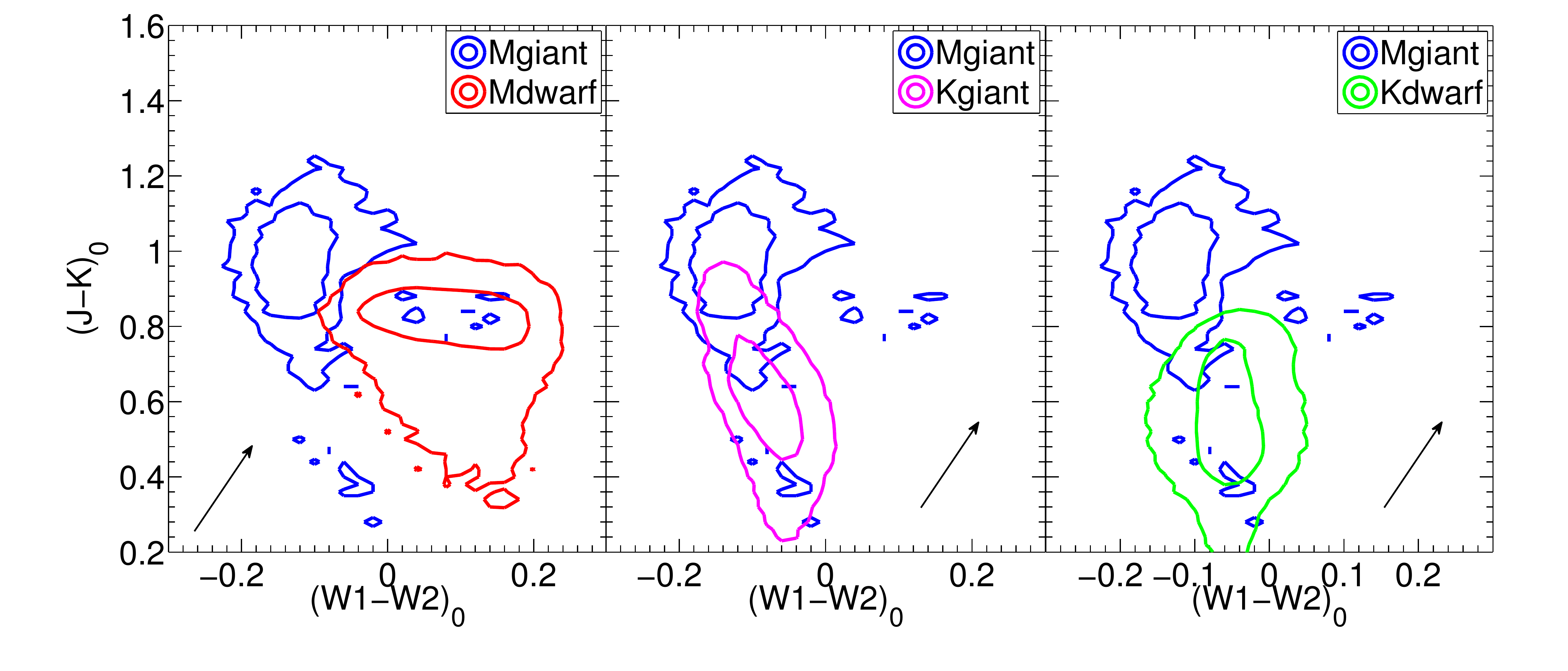}
\includegraphics[width=\columnwidth]{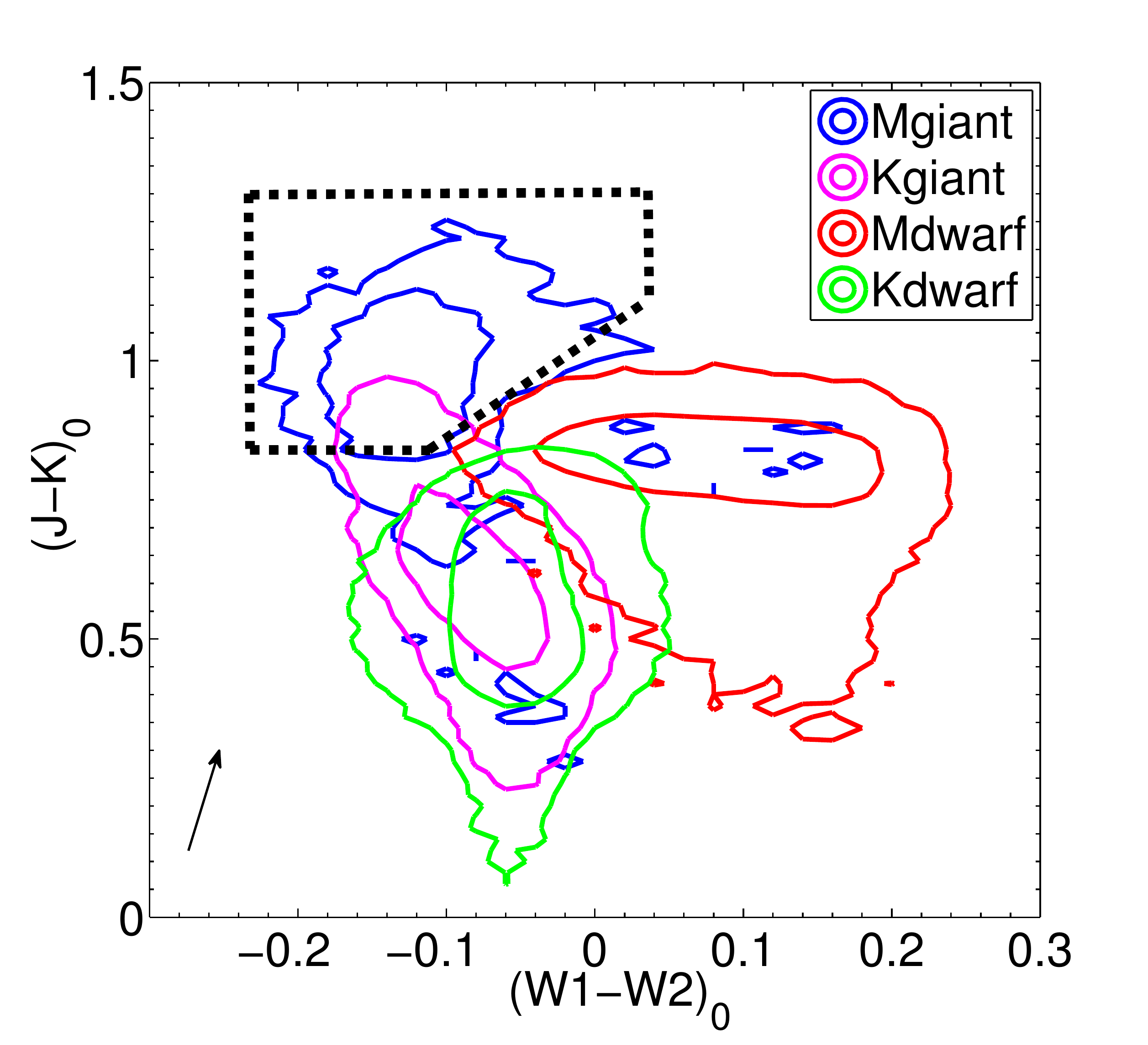}
\caption{2MASS and WISE $(J-K)_0$ vs $(W1-W2)_0$ color-color
  distribution for spectroscopically classified M-giants (blue),
  M-dwarfs (red), K-giants (purple) and K-dwarfs (green). Contours
  correspond to 1- and 2-$\sigma$. 
  The dashed box shows our M-giants selection criteria.
  The black arrow shows an example reddening vector (see Section
  \ref{sec:WISE} for further details).}
\label{KM}
\end{figure}

One issue that is not dealt with in Figure~\ref{wise_CC_distribution}
is the properties of cooler stars, such as K-dwarfs/giants. To address
this we have supplemented our data with a sample
of LAMOST K-type stars from \citet{Liu2014}, which have been 
classified using spectral line features. The distributions in the 
$(J-K)_0$ vs $({\it W1}-{\it W2})_{0}$ plane are shown
in Figure~\ref{KM}.
There is an overlap between late-K and early-M giants, which
is not surprising and most likely due to the \citeauthor{Liu2014}
sample including a number of early M-giants, but reassuringly there is
very little overlap between K-dwarfs and M-giants. This implies that a
2MASS-WISE color selection should be free from significant dwarf
contamination.

Given these findings, we now define the following selection region for
M-giants,
\begin{eqnarray}
-0.23<( {\it W1}- {\it W2})_{0}<0.02,\nonumber\\
0.85<(J-K)_{0}<1.3,\label{eq:selection}\\
(J-K)_{0}>1.45\times({\it W1}- {\it W2})_{0}+1.05.\nonumber
\end{eqnarray}
This has been chosen to minimize contamination while retaining a
reasonable level of completeness. Assuming the spectroscopic
classifications are reliable, we find that there is only 4.9\%
contamination from dwarfs in the color-box given in
Equation~(\ref{eq:selection}). Furthermore, we believe that most of this
contamination is likely due to misclassified spectra; if we remove all
spectra with S/N less than 10, then the contamination from dwarfs
drops to less than 1\%.

Any discussion of contamination and completeness will depend
on the details of the selection-function (e.g. the presence of
color-biases), the sky coverage and the magnitude range under consideration,
with brighter magnitudes containing a higher M-giant/M-dwarf
fraction. This is because at bright magnitudes the volume probed by
M-dwarfs is small and at faint magnitudes the number of M-giants drops
because there are so few at large distances in the halo. For
reference, the magnitude range probed by our sample is $7.63 \la J_0
\la 15.52$.

A similar approach was adopted by \citet{Koposov2014}, although they
chose a narrower selection region in the $(J-K)_0$ vs $(W1-W2)_0$
color-color plane (see equation 1 of their paper). The performance of
their selection is comparable to ours; their completeness is around
10 per cent lower due to the narrower selection, and their
contamination, while still below 5 per cent, is slightly higher than
ours since their selection is closer to the M-dwarf region.

\section{Properties of M-giants}
\label{sec:properties}

\subsection{Metallicity estimation}
\label{sec:metallicity}

In Figure~\ref{KM} we can see that most M-giants are located in the
region $({\it W1}-{\it W2})_{0}<-0.1$, but there is an extension to
redder $({\it W1}-{\it W2})_{0}$ colors. After checking the location
of the stars which populate this extended region we find that most of
them are at high galactic latitudes, indicating that they are likely
to be more metal-poor. This WISE metallicity trend is also predicted
by stellar models, as can been seen (for example) in figure 1 of
\citet{Koposov2014}.

\begin{figure}
\plotone{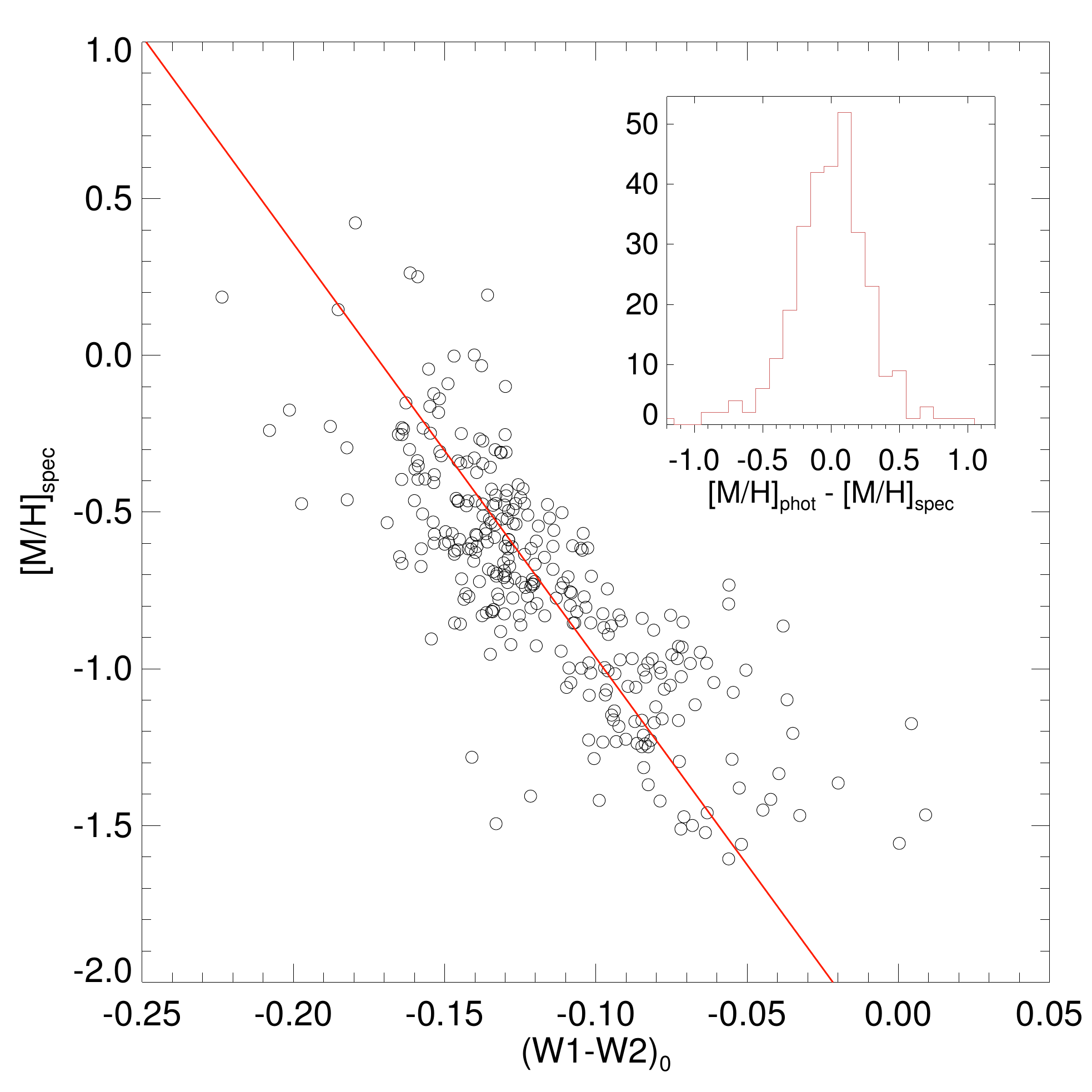}
\caption{The metallicity distribution of APOGEE M-giants vs
$({\it W1}-{\it W2})_{0}$ color. The red line shows the best-fit
linear relationship. The inset histogram shows the scatter in
metallicity about this relation, which has a dispersion of 0.29
dex. Note that the mean error in $\rm [M/H]_{\rm APOGEE}$ is 0.03 dex
and the mean error in $({\it W1}-{\it W2})_{0}$ is 0.03 mag, which
implies that the intrinsic scatter in this relation is exceptionally
small (see Section \ref{sec:metallicity}).}
\label{fig:metal}
\end{figure}

To go one step further, we compare the photometry of our M-giants to
the spectroscopic metallicity obtained using data from the SDSS
project APOGEE \citep{Holtzman2015}.
This survey is taking high S/N and high-resolution (R = 22,500)
NIR spectra, resulting in detailed chemistry and a measurement of 
$\rm [M/H]$ to a precision of better than 0.1 dex. We cross-match
APOGEE DR12 with WISE and 2MASS, using the same photometric quality
cuts as described in Section \ref{sec:WISE} with an additional cut of
$\left|b\right|>30$ deg, and then use our photometric criteria to
select a sample of M-giants.
Figure~\ref{fig:metal} shows the APOGEE metallicities of the resulting
sample of 296 M-giants as a function of $(W1-W2)$ color.
The correlation with $({\it W1}-{\it W2})_{0}$ is very strong and can
be fit with the following linear relation:
\begin{equation}
\rm [M/H]_{phot}=-13.2^{+0.5}_{-0.6}\times({\it W1}-{\it W2})_{0}
 - 2.28^{+0.06}_{-0.07} \hspace{0.15cm} dex,
\label{eq:metallicity}
\end{equation}
taking into account errors on both color and metallicity \citep{Hogg2010}.

As can be seen from the inset of Figure~\ref{fig:metal}, the residual
scatter about this linear relation is small (0.29 dex). What is most
remarkable about this is that the spread is almost entirely due to
photometric uncertainties. The mean photometric error
on $({\it W1}-{\it W2})_{0}$ for these stars is 0.031 mag, which is
larger than the measured spread in $({\it W1}-{\it W2})_{0}$ about the
linear relation (0.022 mag). This implies that the
photometric uncertainties are probably overestimated and, in any case,
the intrinsic dispersion in this relation must be exceptionally
small. If we attempt to fit the relation including a term to represent
the intrinsic spread \citep[Equation 35 of][]{Hogg2010} we find, as
expected, that the best-fit spread is zero.

It should be noted that the APOGEE parameter pipeline encounters
difficulties for very cool giants, with a recommendation that
metallicities for stars cooler than 4000K should be treated with caution
\citep{Holtzman2015}. Around a third of our cross-matched sample are
cooler than this limit, and so to check the effect on our relation we
repeat the fit excluding these stars. The resulting relation is very
similar, although slightly steeper,
\begin{equation}
\rm [M/H]_{phot}=-15.6^{+0.4}_{-0.2}\times({\it W1}-{\it W2})_{0}
 - 2.58^{+0.06}_{-0.04} \hspace{0.15cm} dex.
\label{eq:metallicity2}
\end{equation}
Throughout the paper we adopt the original relation (Equation
\ref{eq:metallicity}) and label these metallicities 
$\rm [M/H]_{phot}$.

Clearly $({\it W1}-{\it W2})_{0}$ is a very good proxy for
$\rm [M/H]$, meaning that this can be used to quantify metallicities
of halo substructures (as we will demonstrate later in Section
\ref{sec:sgr}). Similar work has been carried out by
\citet{Schlaufman2014}, finding that WISE photometry can efficiently
identify metal-poor stars.
 
\begin{figure*}
\centering\includegraphics[width=1.\hsize]{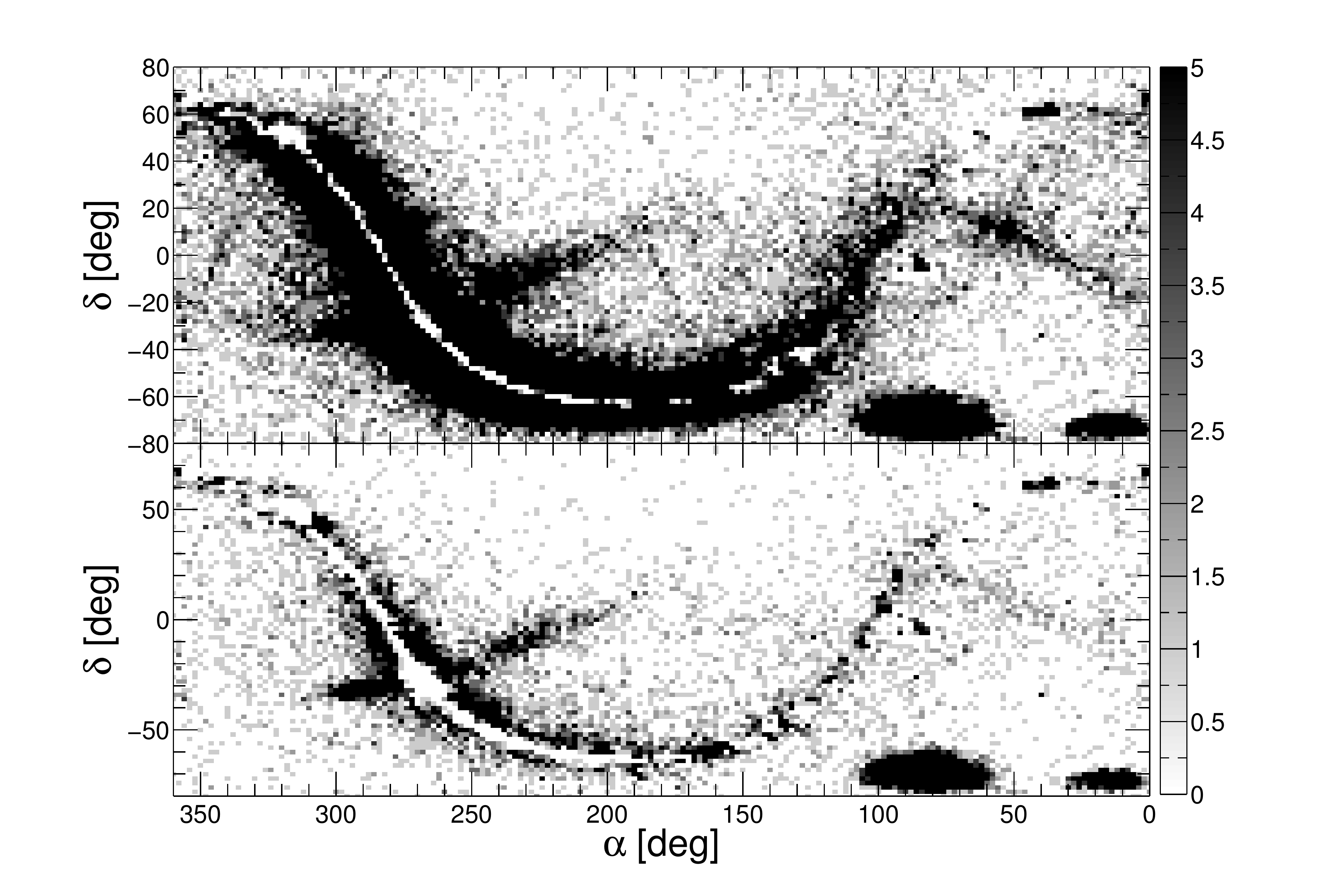}
\caption{Full sky map of M-giants. The upper panel shows all
M-giants, while the bottom panel shows metal-poor M-giants using the
cut $({\it W1-W2})_{0} > -0.13$ mag (corresponding to
$\rm [M/H]_{phot} < -0.56$ dex).
Note that this cut reduces the amount of disc M-giants and also removes
some relatively metal-rich substructures in the halo, such as the
Tri-And system located at RA $\sim$ 30 deg, Dec $\sim$ 30 deg
(see Section \ref{sec:metallicity}).}
\label{fig:fullsky}
\end{figure*}

In Figure~\ref{fig:fullsky} we show the distribution of M-giants on
the sky, using our M-giant selection from Equation
(\ref{eq:selection}) and photometric quality cuts as described in
Section \ref{sec:WISE}. We add a further cut of $J_0>12$, to ensure
that our stars are beyond $\sim10$ kpc, and $10.5<W1_0<13.5$, which is
optimal for detecting the Sgr stream. Using our photometric
metallicity $\rm [M/H]_{phot}$, we show two distributions in this
figure: in the top panel we show all M-giants, while in the bottom we
show the more metal-poor ones (by removing stars with 
$({\it W1-W2})_{0}<-0.13$, corresponding to $\rm [M/H]_{phot} > -0.56$
dex). As can be seen from the lower panel, this
cut reduces the amount of disc M-giants in the sample and also removes
some relatively metal-rich substructures in the halo, such as the
Tri-And system located at RA $\sim$ 30 deg, Dec $\sim$ 30 deg
(\citealt{Majewski2004,rochapinto2004}; see, for
example, Figure 7 of \citealt{Deason2014} for a plot of the
metallicity distribution).

\subsection{Distance estimation}
\label{sec:distance}

\begin{figure*}
\plotone{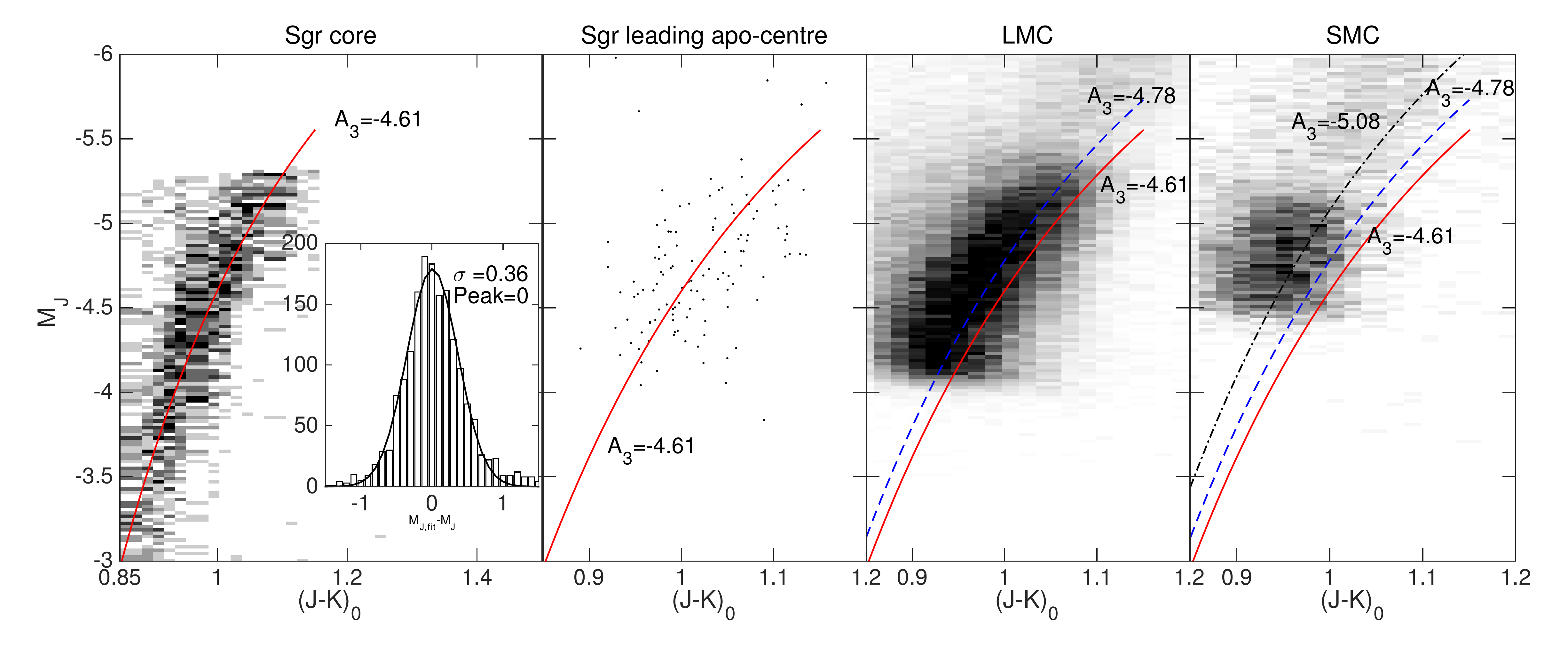}
\caption{Color and $J$-band absolute magnitude relation for four
separate regions: Sgr core, Sgr leading apocenter, the LMC and
SMC.
We calculate $M_J$ assuming all stars lie at the systemic distance of
the respective structure (taking literature values for these
distances; see Section \ref{sec:distance}). The value of $A_3$
corresponds to the offset in absolute magnitude between these
relations (Equation \ref{eq:distance}).
The solid line denotes the best-fit to Sgr core, the
dashed line to the LMC and the dot-dash line to the SMC.}
\label{color_mj}
\end{figure*}

\subsubsection{An empirical distance relation}
\label{sec:distance_empirical}

In Figure~\ref{fig:fullsky}, we can see the Sgr stream, LMC and SMC
very clearly. We can use these structures to estimate an M-giant
color-absolute-magnitude relation, since we already know their
distances. To do this we first calculate the absolute magnitude
($M_J$) for each M-giant in these regions using distance estimates
from the literature, assuming that all stars lie at the systemic
distance of the structure.

We select four regions from the upper panel of
Figure~\ref{fig:fullsky}: a region near the 
core\footnote{We have not included the actual Sgr core in
order to avoid low-latitude regions where extinction and crowding
could affect our analysis.} of Sgr
($285^\circ<\alpha<300^\circ$, $-35^\circ<\delta<-28^\circ$; hereafter
called Sgr core region);
one near the apocenter of the leading tail of Sgr
($214^\circ<\alpha<226^\circ$,$-10^\circ<\delta<0^\circ$); an LMC
region ($60^\circ<\alpha<105^\circ$,$-78^\circ<\delta<-60^\circ$); and
an SMC region ($5^\circ<\alpha<25^\circ$,$-77^\circ<\delta<-70^\circ$).
The distances we adopted for the LMC, SMC, Sgr core and Sgr
leading apocenter are 51 kpc \citep{LMCSMC2006, deGrijs2014}, 61 kpc
\citet{LMCSMC2006}, 29 kpc \citep[calculated using the model
of][]{Law2010}, and 53 kpc \citep{Belokurov2014}, respectively.

Figure~\ref{color_mj} shows $M_J$ vs $(J-K)_0$ relations in our four
regions. As has been seen by many authors \citep[e.g.][]{Sharma2010},
the absolute magnitude for M-giants varies approximately linearly with
$(J-K)_0$ color. However, it is not precisely linear, so we choose to
fit these data using the following power-law relation,
\begin{equation}
\label{eq:distance}
M_{J,\rm fit} = A_1 \left[\left(J-K\right)_0^{A_2}-1\right]+A_3,
\end{equation}
where $A_3$ corresponds to the value of $M_J$ at $(J-K)_0=1$ mag.
We do this by slicing the data into $M_J$ segments and fitting each
of these with two Gaussians (one for the system and one for the
background). We then fit the centroids of the former Gaussians with
the above polynomial relation.

We start with the Sgr core region, which covers the largest magnitude
range (owing to the fact that it is the closest of the three
structures). The parameters are presented in Table
\ref{tab:distance} and should be valid for the color range that we
probe, i.e. $0.9 \la (J-K)_0 \la 1.1$. The inset in the first panel of
Figure ~\ref{color_mj} shows the scatter around the $M_{J,\rm fit}$
relationship. The dispersion is 0.36 mag, which corresponds to a
distance uncertainty of $20\%$. Note that this uncertainty doesn't
take into account population effects (which we discuss next) and so
this estimate can be considered as a lower-limit on the actual uncertainty.

\begin{figure}
\plotone{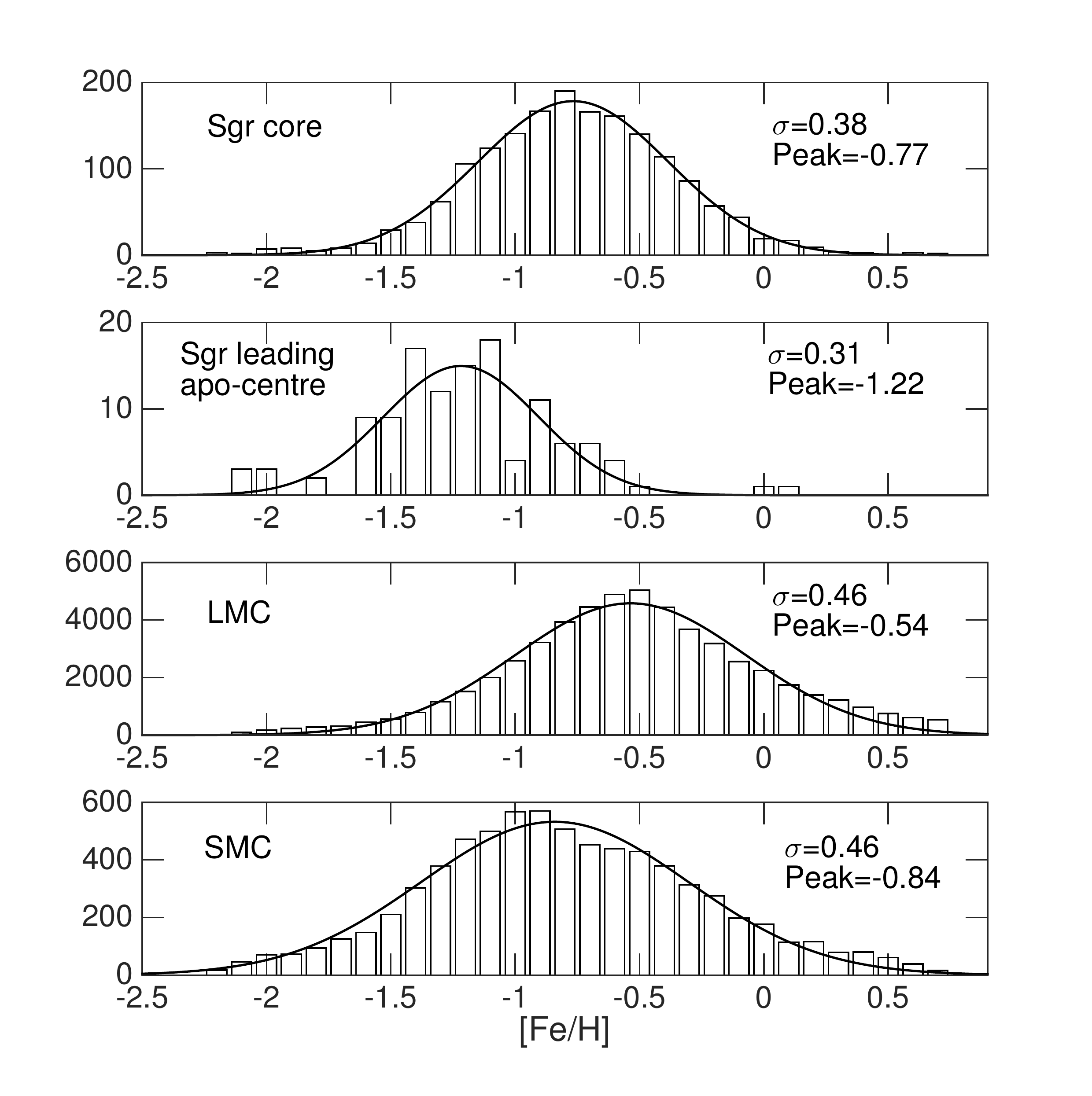}
\caption{Photometric $\rm [Fe/H]$ distribution in Sgr core, Sgr
apocenter, LMC and SMC region, as derived from Equation
(\ref{eq:metallicity}).}
\label{metal_dist}
\end{figure}

In the second panel of Figure~\ref{color_mj} we
compare the Sgr core fit to data from the leading tail's apocenter,
finding very good agreement. The final two panels of this figure show
the derived relation for the LMC and SMC, where we have kept
parameters $A_1$ and $A_2$ fixed to match the Sgr core values. As
expected the color-absolute-magnitude relation varies between Sgr and
the LMC and SMC, owing to the different metallicities and
star-formation histories of these systems.
Figure~\ref{metal_dist} shows the metallicity distribution function
(MDF) in these four regions, using the photometric metallicity
relation presented in Equation (\ref{eq:metallicity}).
It is interesting to note that although the MDF for the Sgr core and
SMC are quite similar, their $M_{J,\rm fit}$ relations vary
significantly. In fact, despite having a significant shift in mean
metallicity compared to Sgr core, the LMC has a much better agreement
in terms of $M_{J,\rm fit}$. These discrepancies are likely caused by
the different star-formation histories in these systems. This can also
be seen when one considers the difference between the Sgr core and
leading apocenter regions, which have very similar $M_{J,\rm fit}$
relations yet clearly discrepant MDFs. This is most likely a
consequence of the fact that the leading apocenter material is
composed of stars stripped from the outer parts of the infalling
dwarf, while the core material is from the younger populations in the
center. So the shift in $M_{J,\rm fit}$ from the MDF offset is
probably canceled by a corresponding shift coming from the age offset.
We return to the issue of Sgr in Section \ref{sec:sgr}.

We performed the following simple test to check whether internal
extinction in these systems is likely to be affecting our derived
$M_{J,\rm fit}$ relations.
We took two LMC samples, one consisting of the entire LMC region
($60^\circ<\alpha<105^\circ$, $-78^\circ<\delta<-60^\circ$) and one
covering the same area but excluding the central region
($70^\circ<\alpha<90^\circ$, $-72^\circ<\delta<-68^\circ$) where
internal extinction will be most significant. We then performed the
same $M_{J,\rm fit}$ fitting to these two regions and found that there
was no clear shift in the relations, confirming that internal
extinction is unlikely to bias our results.

\begin{table}
\begin{center}
\caption{The parameters for our color-absolute-magnitude relation
($M_{J,\rm fit}$), as given in Equation (\ref{eq:distance}). For the Sgr
core region all three parameters are left free. For the other two
regions only $A_3$ is free and the remaining parameters are fixed to
the values found for the Sgr core. These relations should be valid for
$(J-K)_{0}$ in the range 0.9 -- 1.1 mag.}
\label{tab:distance}
\begin{tabular}{lccc}
\hline
$region$ & $A_1$ & $A_2$ & $A_3$ \\
\hline
Sgr core & 3.12 & -2.6 & -4.61\\
LMC & -- & -- & -4.78\\
SMC & -- & -- & -5.08\\
\hline
\end{tabular}
\end{center}
\end{table}

\begin{figure}
\plotone{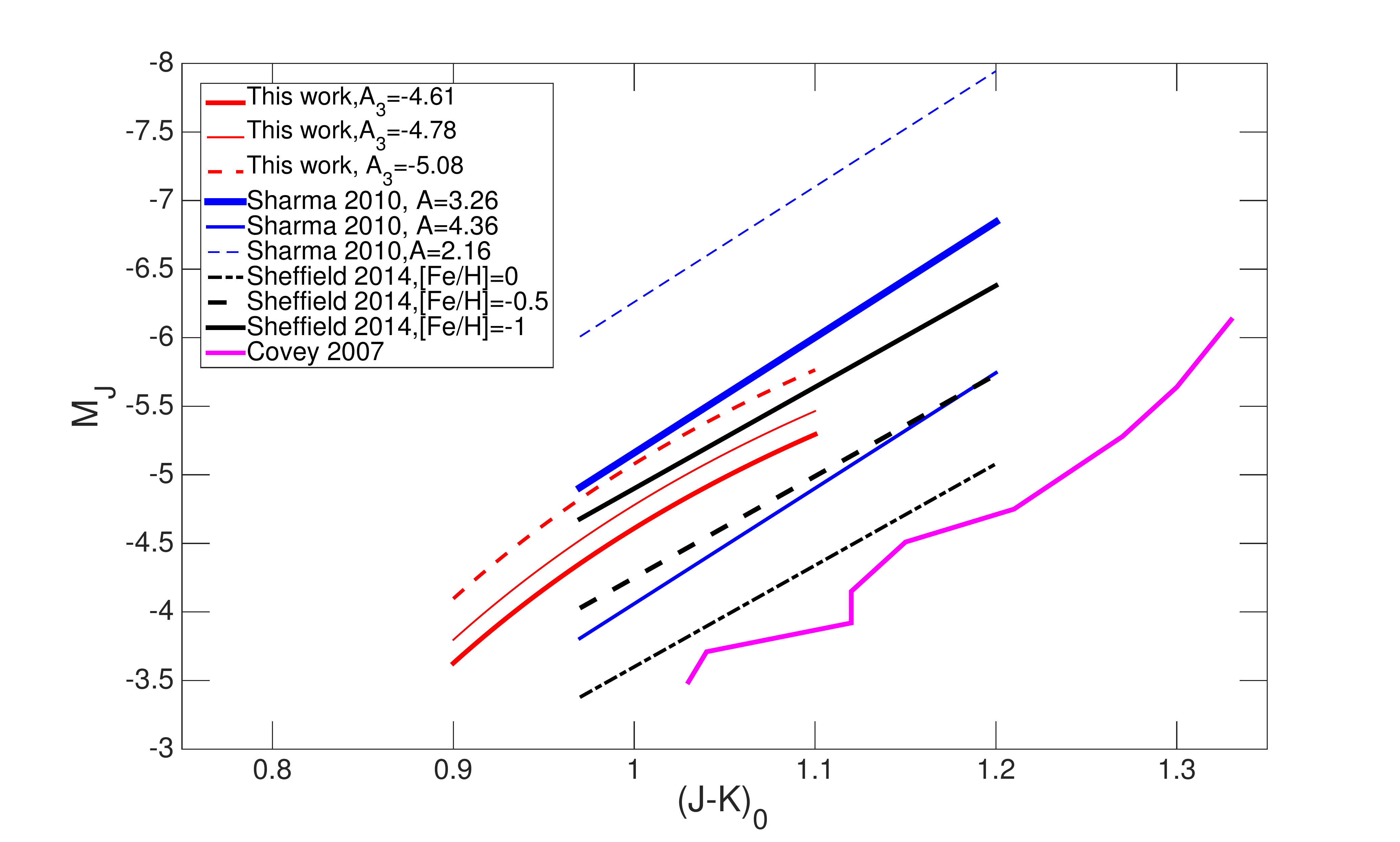}
\caption{Comparison of our empirical distance relation to various
literature sources. The red lines correspond to our new relations,
where the top, middle and bottom lines correspond to relations
based on the SMC, LMC and Sgr, respectively.
Our relations are only valid for the range $0.9 \la (J-K)_0 \la 1.1$,
while the \citeauthor{Sharma2010} and \citeauthor{Sheffield2014} ones
are valid for $0.97 \la (J-K)_0 \la 1.2$.}
\label{dist_compare}
\end{figure}

In Figure~\ref{dist_compare} we compare our $M_{J,\rm fit}$ relation
to some of those in the literature.
The two main references we use are \citet{Sharma2010} and
\citet{Sheffield2014}. The former paper uses a single linear relation
calibrated to match a range of stellar models with ages and
metallicities consistent with simulated stellar halos
\citep{Bullock2005}. \citet{Sheffield2014} generalized this to include
a metallicity dependence, again using stellar models.
For completeness we also include
the relation from \citet{Covey2007}, which the authors admit is not
especially robust (this relation is itself based on
\citealt{Pickles1998}, who use solar-age and solar-abundance stellar
models).
Note that our method differs from these in that it is an empirical
relation, rather than relying on stellar models.
We can see that our three relations (Sgr, LMC and SMC) are
consistent with the $\rm [Fe/H]=0$ relation from
\citet{Sheffield2014}, even though these three systems are all more
metal-poor than this value. In particular our Sgr relation, which
provides a good match to the leading apocenter where $\rm [Fe/H] \la
-1$, is offset from \citeauthor{Sheffield2014}'s $\rm [Fe/H]=-1$
relation by around 1 mag. As discussed above, this is may be due to
the age dependence of these relations. We conclude that a simple
relation based on metallicity might overlook some subtleties and hence
not be ideal for all purposes.

\subsubsection{Metallicity dependence and age-metallicity trends in
the LMC and Sagittarius}

The location of the red giant branch in color-magnitude space is
well-known to be dependent on both age and metallicity. Given the
metallicity relation found in the previous section, the question
naturally arises whether we can detect metallicity gradients within
our distance relation.

We first test the core region of Sgr, where we have the most coverage
of the color-magnitude plane. We repeat the same procedure as above,
assuming all stars lie at a distance of 29 kpc, but for two samples:
one with metallicity around 0.5 dex greater than the mean and one
around 0.5 dex smaller ($-0.42 < [M/H] < 0.00$ dex and $-1.50 < [M/H]
< -1.12$ dex, respectively, where the mean photometric metallicity is
$-0.76$ dex). In order to clean the sample, in each band we reject the
stars with 10\% largest photometric errors. The two samples have
around 250 stars each.

\begin{figure*}
\plottwo{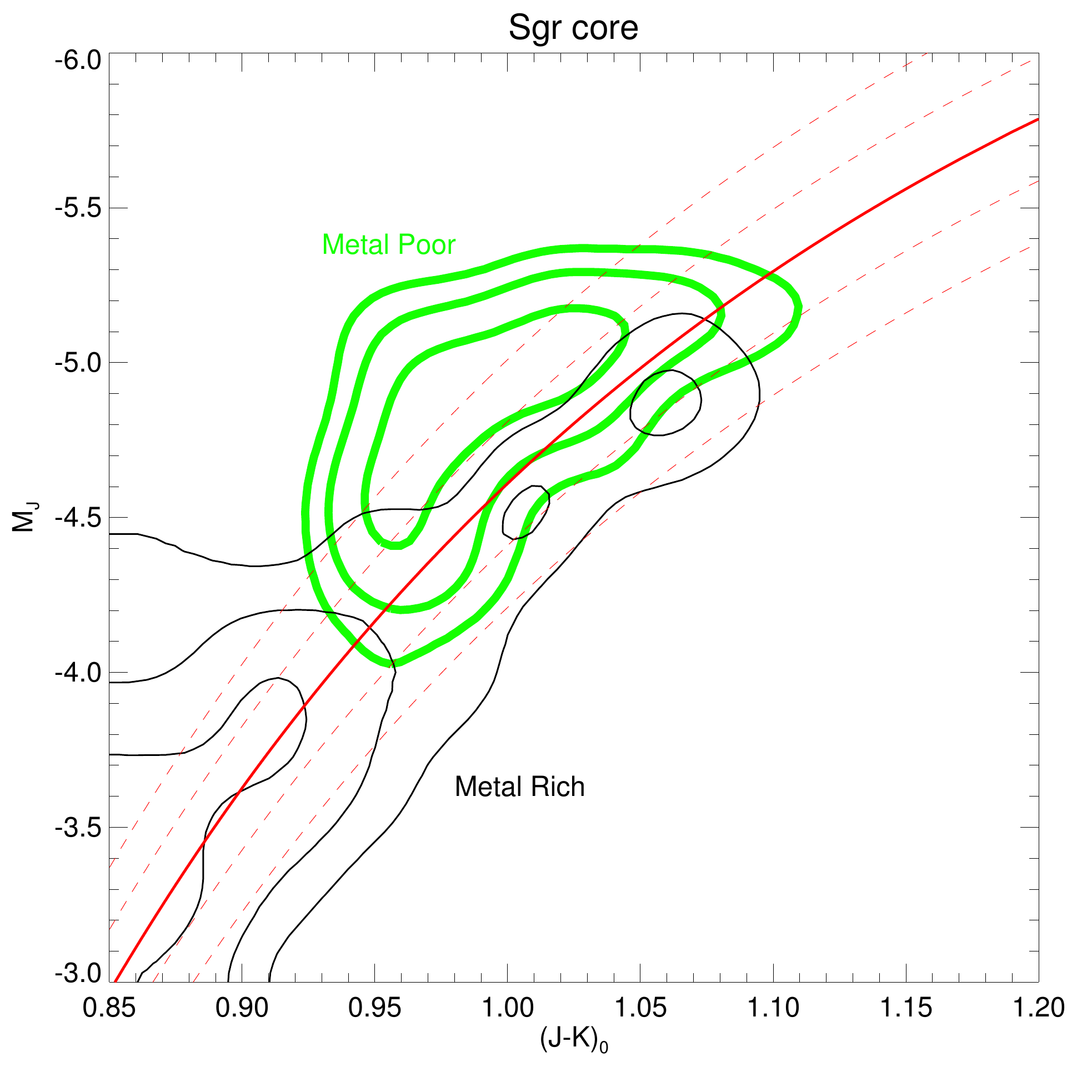}{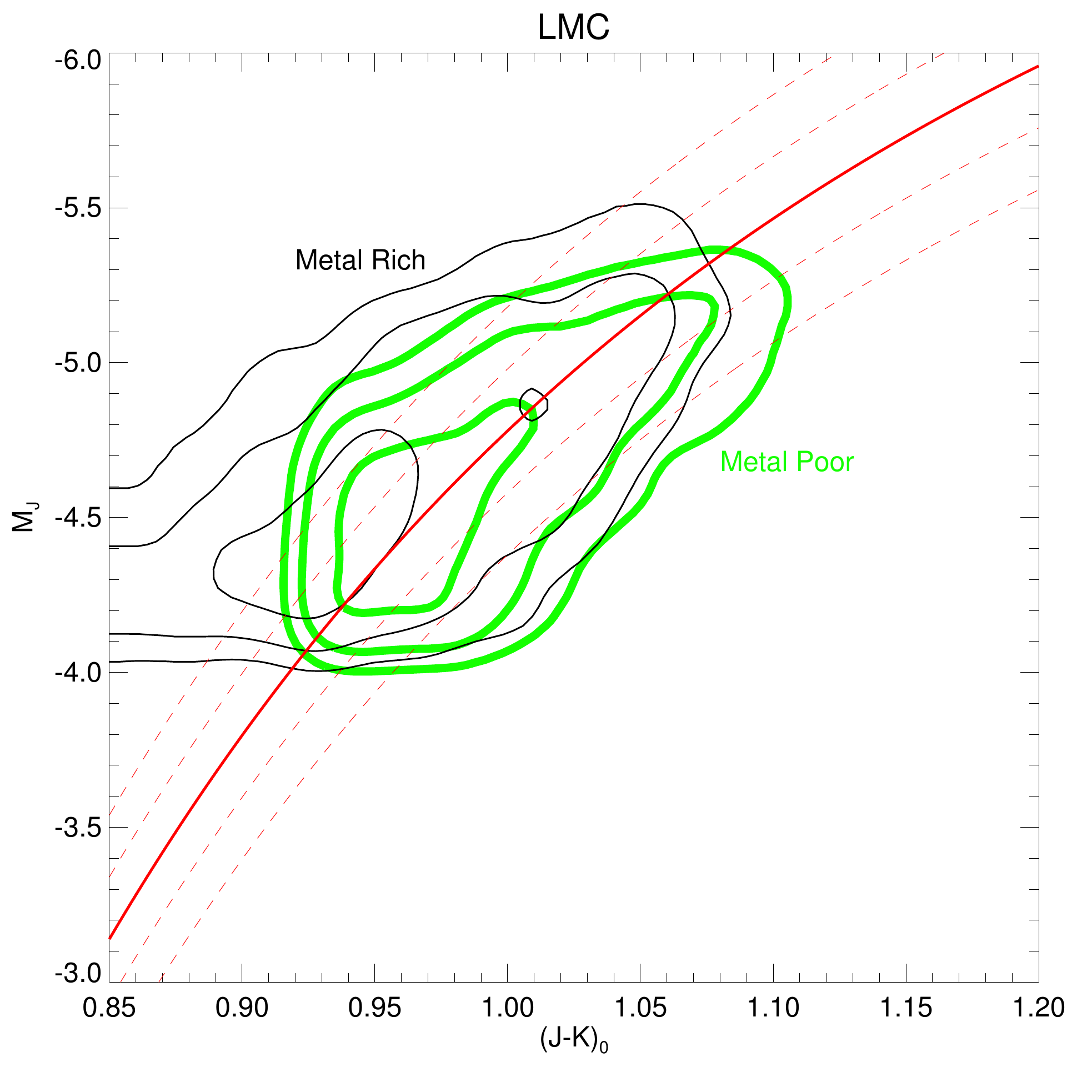}
\caption{The dependence of our derived $M_J$ vs $(J-K)_0$ relation on
metallicity. The thick contours show a sample of stars which are 0.5
dex more metal-poor than the overall population, while the thin
contours show a sample which is 0.5 dex more metal-rich. Contours
correspond to 0.3, 0.5 and 0.8 times the peak value. The solid line
denotes the fiducial relation derived in Section
\ref{sec:distance_empirical}, and the dashed lines correspond to
$\pm0.2$ and $\pm0.4$ mag in $M_J$.}
\label{fig:dist_met}
\end{figure*}

The resulting distributions of $M_J$ vs $(J-K)_0$ are shown in
Figure~\ref{fig:dist_met}. The fiducial relation derived for the whole
population (Section \ref{sec:distance_empirical}) is shown by the
solid red line. As can be seen, especially for the region 
$0.95 < (J-K)_0 < 1.10$ mag, there is a systematic shift with the
metal-rich population lying below the relation and metal-poor one
above. This matches the expectation from stellar models, where
for a given age metal-poor populations are bluer. The dashed lines in
this Figure correspond to the fiducial relation shifted vertically by
$\pm0.2$ and $\pm0.4$ mag, which means that the shift with metallicity
for Sgr is around 0.3 to 0.6 ${\rm mag.dex}^{-1}$.

It can also be seen that the metal-rich population extends to smaller
$(J-K)_0$ colors. This is due to our diagonal color-cut, which
means that at redder $(W1-W2)_0$ colors the M-giant selection box does
not extend as far in $(J-K)_0$ (see Figure~\ref{KM}).

We repeat this exercise for the LMC, where we have the largest number
of M-giants. We again reject lower-quality photometry, this time
removing the stars with 20\% largest photometric errors in each band.
Our two metallicity ranges were again chosen to be 0.5 dex above and
below the mean value ($-0.20 < [M/H] < -0.05$ dex and $-1.20 < [M/H] <
-1.07$ dex, respectively, where the mean photometric metallicity is
$-0.63$ dex). Each sample has around 2,500 stars.

The resulting distributions, shown in Figure~\ref{fig:dist_met}, are
significantly different from those of Sgr. There is no clear trend
with metallicity and one might even claim that the metal-rich
population is bluer than the metal-poor one. The contrasting behavior
for the LMC and Sgr can be understood when one considers the
difference between the star formation histories of these two
systems. We know that the LMC retains gas and has an extended star
formation history. As a consequence, the metal-rich populations are
likely to be significantly younger than the metal-poor ones, which can
lead to the $M_J$ dependence on metallicity to be canceled out. This
is because the red giant branch for younger populations is generally
bluer than older populations. However, the star formation history
of Sgr is likely to be less extended than the LMC, as its infall into
the Milky Way will have removed most of the gas and truncated star
formation (see, for example, \citealt{deBoer2015}).

\subsubsection{Test case: The tilt of the LMC}

Although the distance uncertainty for an individual M-giant is
relatively large (around 20\% before considering population effects),
for structures like the LMC we can average large numbers of stars to
determine the distance to relatively high precision. We therefore test
the performance of our relation by addressing the tilt in the LMC
disc, which has been established by previous authors
\citep[e.g.][and references therein]{Subramanian2013}.

\begin{figure}
\plotone{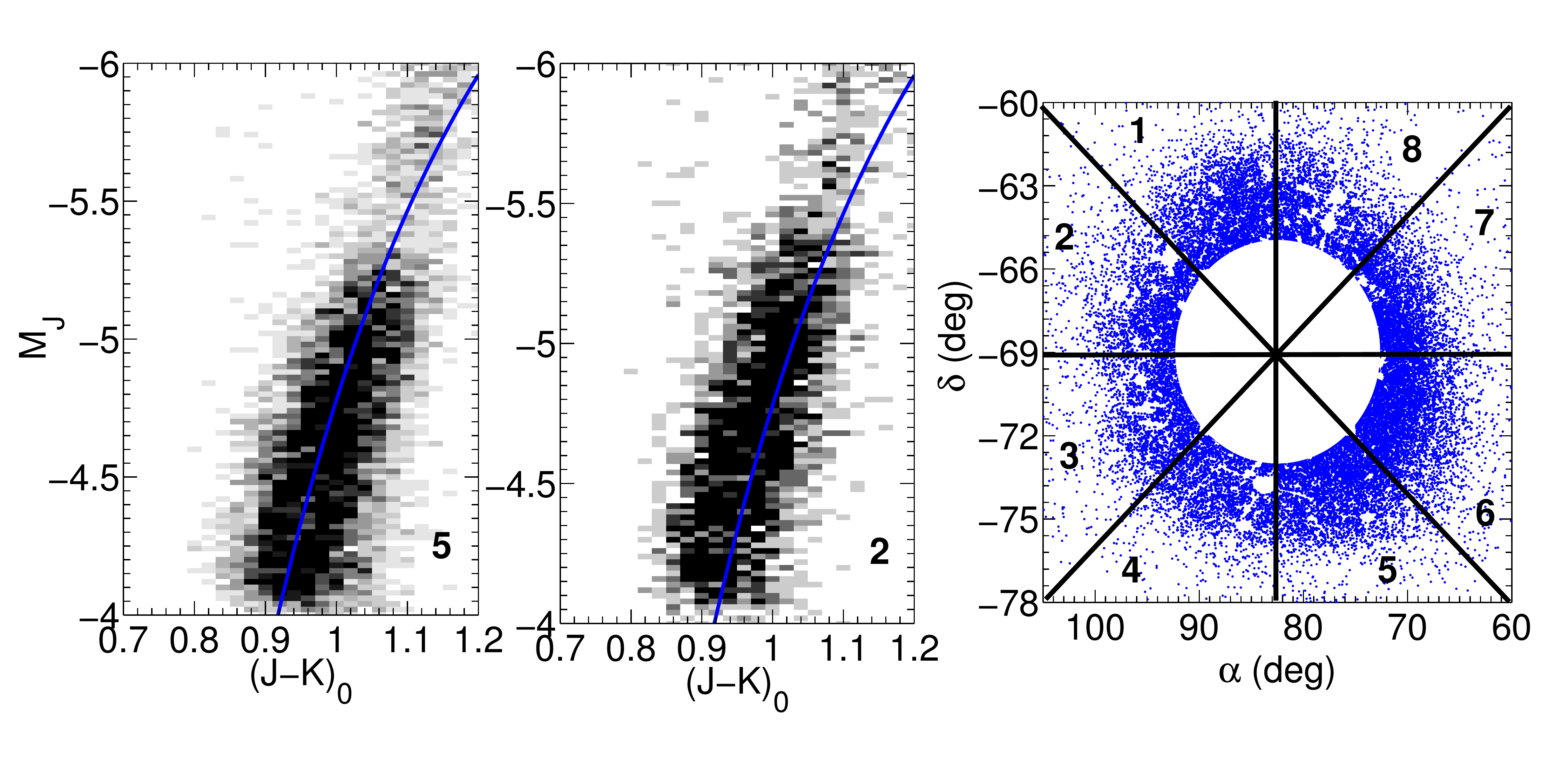}
\caption{Our determination of the tilt of the LMC. We divide our data
into eight azimuthal segments (right panel) and for each of these plot
the color-absolute-magnitude distribution assuming a systemic
distance of 51 kpc. The left and middle panels show segments 2 and
5, which are the closest and most distant segments, respectively. The
solid line denotes the fiducial relation calculated in Table
\ref{tab:distance}. There are clear distance offsets in these
regions, reflecting the fact that the LMC is tilted to our
line-of-sight. Note that we have excluded the inner regions as we do
not expect these will exhibit significant tilt.}
\label{LMC}
\end{figure}

We do this by calculating $M_J$ for each of the stars, assuming a
distance of 51 kpc, and compare the distribution of magnitudes to the
fiducial relation calculated above (see Table \ref{tab:distance}). The
tilt means that stars on the near-side will be on-average
brighter, while those on the far-side will be on-average fainter.
Since we expect such variations to be correlated with azimuth, we 
divide up the data into eight segments as shown in the right panel of
Figure~\ref{LMC}.
The central regions will exhibit negligible tilt, so we exclude these
using an ellipse with major axis of 10 deg and minor axis of 4 deg,
centered on ($\alpha$, $\delta$)=($83^\circ$, $-69^\circ$).
The left and middle panels show the color-absolute-magnitude relation
for two of these segments, compared to the fiducial relation.
There are clear magnitude offsets in some of these regions, with Segment
2 (middle panel) hosting stars that are on-average brighter than the
fiducial relation, while Segment 5 (left panel) shows the opposite
behavior. To quantify this discrepancy, we calculate the mean 
offset from the fiducial relation for each numbered segment by fitting
the distribution with a Gaussian function. The results are presented
in Table~\ref{tab:tilted}. From this we can see
that the offsets exhibits a clear sinusoidal behavior with azimuth, as
expected if the LMC is tilted to our line-of-sight. By fitting a sine
curve to these offsets (excluding Segment 7 which has an unusually
broad M-giant branch and does not match the overall trend), we find
that the position angle of the line of nodes is $128.5^\circ$, where
the angle is measured Eastwards from the North. This compares
favorably to \citet{Subramanian2013}, who found an angle of
$141.5^\circ$.

We can also investigate the inclination angle, but this is slightly
harder to measure as it is also a function of the distance of the
stars from the center of the LMC. The amplitude of the offsets in
Table \ref{tab:tilted} is around 0.11 mag and the separation of these
stars is around 10 deg from the center of the LMC. This corresponds to
an inclination of around 16 deg, which is not too dissimilar to the
value of 25.7 deg found by \citet{Subramanian2013}.

\begin{table}
\begin{center}
\caption{The values of the mean offset in J-band magnitude as a
function of azimuth for 8 segments around the LMC. The segment number
corresponds to those in Figure~\ref{LMC}.}
\label{tab:tilted}
\begin{tabular}{lccc}
\hline
Segment & Mean offset\\
& (mag)\\
\hline
$1$ & -0.151\\
$2$ & -0.184\\
$3$ & -0.101\\
$4$ &  0.004\\
$5$ &  0.029\\
$6$ &  0.021\\
$7$ & -0.192\\
$8$ & -0.121\\
\hline
\end{tabular}
\end{center}
\end{table}

\section{Mapping the Sagittarius stream}
\label{sec:sgr}

In Figure \ref{fig:fullsky} we have shown the distribution of M-giants
across the entire sky. The most prominent features in this figure are
the LMC and SMC (located in the bottom-right corner of each panel),
but in addition the Sgr stream is clearly visible, extending from the
core of the dwarf galaxy located behind the Galactic bulge (lying at
${\rm RA = 284}$ deg and ${\rm Dec = -30}$ deg). This map is qualitatively
similar to that obtained from 2MASS M-giants \citep{Majewski2003}, but
as we have shown above our version will have less M-dwarf
contamination.\footnote{Our map is very similar to that of
\citet[][right panel of Figure 2]{Koposov2014}, owing to the fact that
our 2MASS \& WISE selection cuts are similar to theirs (see Section
\ref{sec:2MASS_WISE}).}
There are further stream-like detections (e.g. at
${\rm RA, Dec} = 180, -10$ deg or $90, -20$ deg), but these are
most-likely artifacts due to the WISE scanning law.
We present a table of all M-giants from this map in Appendix
\ref{sec:online_table}.

\begin{figure}
\plotone{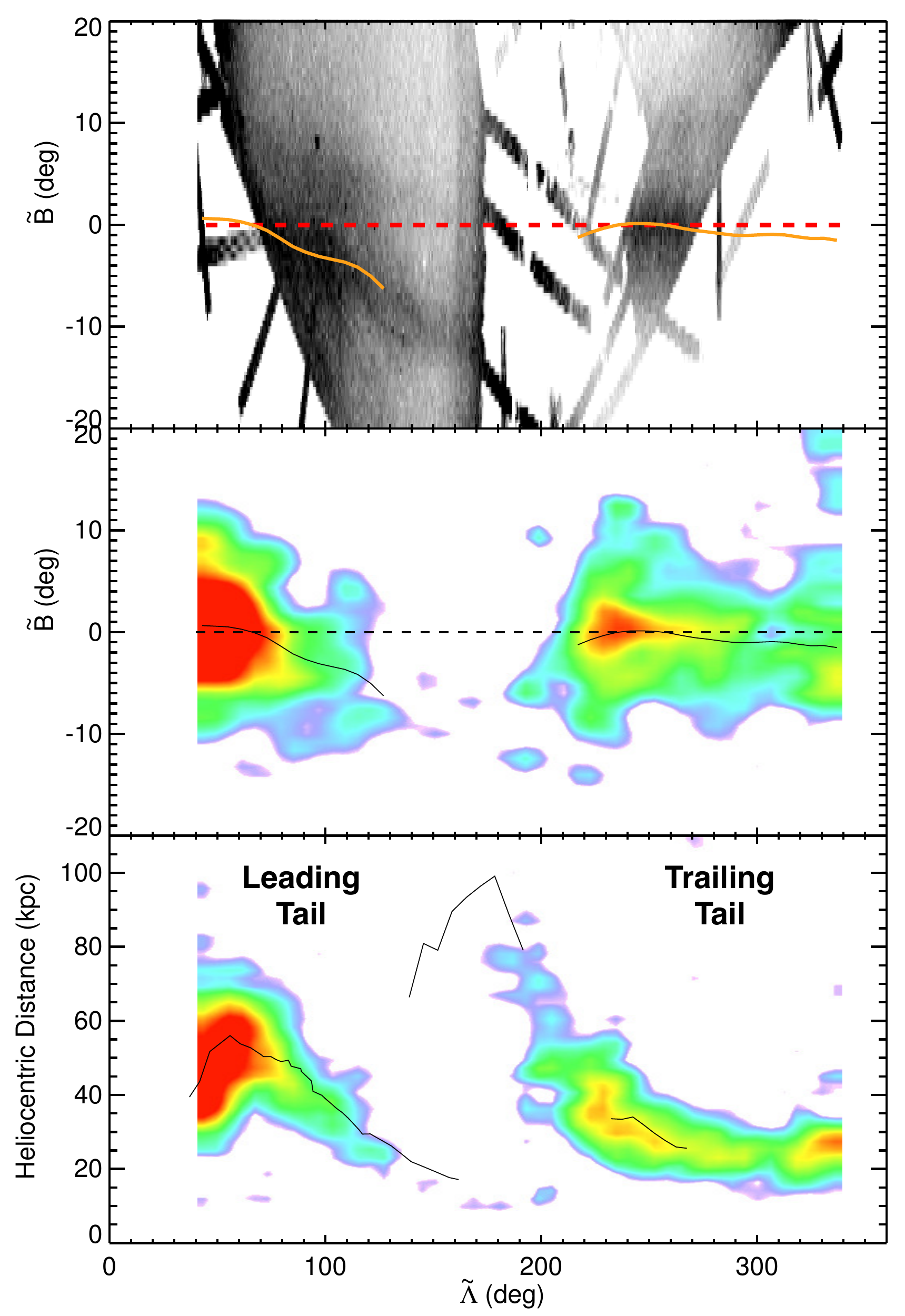}
\caption{
Sagittarius stream tomography and distance.
The middle panel shows the density of our WISE-selected M-giants in
the Sgr coordinate system 
\citep[see Appendix of][]{Belokurov2014}. In this system the Sgr dwarf
is located at (0,0) and 
its orbit moves in the direction of increasing $\tilde{\Lambda}$. We
have only included stars which have distances within 10 kpc of the stream.
The centroid of the stream is denoted by the solid line.
The top panel shows the equivalent plot for MSTO stars \citep[taken
from][]{Koposov2012} and the solid line is the same as the middle panel.
The lower panel shows heliocentric distance measurements of our
WISE-selected M-giants for $-10 < \tilde{B} < 10$ deg. Solid lines
show detections from \citet{Belokurov2014}, consisting of subgiant,
red giant branch and blue horizontal branch detections.}
\label{fig:sgr_allsky}
\end{figure}

\subsection{Distances to the stream}

Given the relations obtained in the previous section, we now proceed
to analyze the distribution of Sgr M-giants. We start off by looking
at the heliocentric distances to the stream. This is shown in the
bottom panel of Figure \ref{fig:sgr_allsky}, where we have plotted all
M-giants obtained using the selection criteria defined in Equation
(\ref{eq:selection}) but restricted ourselves to only those stars which
lie within 10 deg of the Sgr orbital plane
(defined following the equations in the Appendix of
\citealt{Belokurov2014}; note that this coordinate system is related
to the \citealt{Majewski2003} system through
$\tilde{\Lambda}=360-\Lambda$ and $\tilde{B} = -B$).
Distances have been calculated using the Sgr relation given in Section
\ref{sec:distance}. Background subtraction has been performed by
averaging two fields at $10 < \tilde{B} < 20$ deg and $-20 < \tilde{B}
< -10$ deg. We have omitted the two regions closest to the bulge
($\tilde{\Lambda} < 40$ deg and $\tilde{\Lambda} >320$ deg) as
confusion means it is difficult to reliably identify the Sgr stream.

This plot clearly shows the leading stream reaching apocenter at
around 50 -- 60 kpc at $\tilde{\Lambda} = 50$ -- 60 deg, and the
trailing stream at $\tilde{\Lambda} >$ 220 deg. We see good agreement
with literature distances \citep[shown by the solid black lines in
this panel; taken from][]{Belokurov2014}, although there is a possible
discrepancy in the gradient for the leading debris around
$\tilde{\Lambda} \ga 100$ deg, which may be due to population effects
affecting our distance relation. This may occur if the material
more-distant from the core is more metal-poor or older, having been
stripped earlier. We return to the issue of metallicities later in
this Section. It is worthwhile to note that although these literature
distances come from a variety of tracers (blue horizontal branch
stars, red clump stars, subgiant branch stars and now M-giants), the
agreement is remarkably consistent.

There has been recent controversy over the potential detection of the
apocenter of the trailing tail of the Sgr stream 
by \citet{Belokurov2014}, based on BHBs and kinematics of giant stars
in SDSS (see also \citealt{Newberg2003} and
\citealt{Drake2013}). This detection is shown with the distant black
line in the bottom panel of Figure \ref{fig:sgr_allsky}. Recent
evidence to support this claim comes from \citet{Koposov2014}, where
they take velocities of M-giants around $\tilde{\Lambda} \sim
200$--230 deg and find good agreement between the existing trailing
debris and the proposed apocenter detection.
Making a detection of this connecting material is difficult due to
the fact that at this point the stream passes through the disc in the
Galactic anti-center. However, given the purity of the giant sample
from our WISE selection, we are also able to support the hypothesis
that the distant detection is the apocenter of the trailing
tail. Although the WISE data is not deep enough to reach the apocenter
itself, in the lower panel of Figure \ref{fig:sgr_allsky} we clearly
detect a bridge of stars connecting the well-studied trailing material
at $\tilde{\Lambda} > 210$ deg with the apocenter detection at
$\tilde{\Lambda} < 190$ deg. If one takes all M-giants in this region
($180 < \tilde{\Lambda} < 210$ deg and $55<{\rm D_{helio}}<80$ kpc),
there is an overdensity around the plane of the stream (with
$\tilde{B}\sim5$ deg).

\citet{Belokurov2014} also argued that the trailing stream may extend
even further, connecting to an over-density located around
$\tilde{\Lambda} \sim 110$ deg (which is sometimes referred to as
Branch C; see \citealt{Belokurov2006}, \citealt{Correnti2010}).
We also find a tentative detection of this overdensity (with around
10--15 M-giants; this is the weak spur at $\tilde{\Lambda} = 115$
deg and at a distance of 45 kpc in Figure \ref{fig:sgr_allsky}), implying
that its existence is robust as it has been detected in samples of
subgiants, red clump stars and now M-giants. It is indeed clustered
around the Sgr orbital plane, but appears to be relatively broad (with
$-15\la\tilde{B}\la5$ deg). One may expect the stream to be more
diffuse this far from the core. However, until one can detect material
connecting Branch C to the apocenter detection (which is not evident
in our data), then its nature remains up for debate.

\subsection{Orbital plane}

In the middle panel of Figure \ref{fig:sgr_allsky} we have plotted the
distribution of Sgr stream stars on the sky, rotated into the Sgr
coordinate system \citep{Belokurov2014}. In this system Sgr lies at
(0,0) and its orbit moves in the direction of increasing
$\tilde{\Lambda}$. To construct this figure we have only included
M-giants within 10 kpc of the stream (as given by the black lines in
the lower panel, where we have linearly extrapolated for the regions
with no detections). We have also incorporated a background
subtraction, averaging two regions either side of this range
(i.e. $20\la\tilde{B}\la30$ deg and $-30\la\tilde{B}\la-20$ deg) and
interpolating between them.

As mentioned above, it is difficult to determine the behavior
towards the anti-center ($\tilde{\Lambda} \sim 180$ deg), but apart
from that the stream is clearly detected and lies close to the adopted
stream plane (i.e. $\tilde{B} = 0$). We have fit the cross-stream
profile with a Gaussian model in order to trace the centroid of the
density distribution and this is shown by the two solid lines. The
trailing tail lies almost precisely on the orbital plane. The leading
stream, however, shows clear deviation from the plane and crosses at a
shallow angle. This offset between the orbital planes of the leading
and trailing tails is not new; for example \citet{Johnston2005} did
this with 2MASS M-giants, finding an offset of around 5--10 degrees in the
orientation of the two poles. More recent studies have been carried
out by \citet{Newby2013} and \citet{Belokurov2014}.

In order to interpret this plot, we now compare this to the
behavior of main-sequence turn-off (MSTO) stars in this region. Many
authors have looked at the distribution of MSTO stars and we have
included a map of MSTO density in the top panel of Figure
\ref{fig:sgr_allsky} (taken\footnote{Note that there is an error in
the axis labeling of Figure 1 in \citet{Koposov2012}, as the $B$ axis
has the wrong sign. See Figure 2 of \citet{Belokurov2014} for a
correct version.} from \citealt{Koposov2012}).
The MSTO stars exhibit the well known `bifurcation' in the leading
stream, splitting into two separate branches bisected by $\tilde{B} =
0$ with the faint companion stream at $\tilde{B} \sim 2$
deg. \citet{Koposov2012} also detect a bifurcation in MSTO stars in
the trailing stream, with the bifurcation being harder to detect and
consisting of a faint wing to the distribution 
(around $-15<\tilde{B}<5$ deg; see also \citealt{Slater2013}).

Our distributions (also shown by the solid lines in the top panel)
show good agreement with the bright MSTO stream in both the
leading and trailing tails, but with no evidence for a bifurcation.
The location where the bifurcation is clearest in MSTO stars is
around $100<\tilde{\Lambda}<160$ deg, but here the density of M-giants
is low and so it is hard to detect even the bright stream.
Therefore this non-detection may be because there are insufficient
stars to sample the faint secondary stream, or this may be due to
physical differences between the two streams.
There are two main hypotheses for the origin of the faint stream:
either this is material from an earlier wrap of the stream
\citep[e.g.][]{deBoer2015}, or this is from a smaller companion system
to the main progenitor \citep[e.g.][]{Koposov2012}. Both of these 
scenarios are consistent with our non-detection of the faint stream,
if one argues that the faint stream should posses fewer M-giants as it
is older and more metal-poor than the bright stream (because
the faint stream contains material stripped from the outer parts of
the progenitor where there is less metal-enrichment, or because it is
simply from a smaller system with less metal-enrichment).
One final point to note is that \citet{Koposov2012} claimed to detect
a bifurcation in the distribution of 2MASS-selected M-giants in the
trailing tail, with the faint stream lying at $5<\tilde{B}<10$ deg for
$240<\tilde{\Lambda}<270$ deg. We however are unable to confirm this
with our WISE-selected sample.

Our detection shows very good agreement with the bright stream,
although the leading tail appears to be misaligned by a couple of
degrees. This could be due to statistical fluctuations, especially if
a few M-giants from the faint stream were biasing the centroid
fits. The other explanation is that this offset is due to population
effects, as one would not necessarily expect that the M-giants and
MSTO stars to share the same radial profile in the progenitor
system. Rather we expect that the M-giants, which are rare in halo
populations, should be more tightly bound and centrally concentrated
than the MSTO stars, and this may lead to the two populations being
stripped differently and hence not lying on identical orbital
planes. We defer a detailed analysis of the alignment of the orbital
planes to a further study. An in-depth discussion of the alignment and
precession of the stream has been presented in \citet{Belokurov2014},
which compares the orbital plane of the MSTO and red clump stars to
that of the M-giants as measured by \citet{Johnston2005}. They also
find that the M-giants lie close to the plane of the bright stream,
with the orbital poles being a few degrees offset for the leading
stream and almost identical for the trailing stream.

\subsection{Metallicity gradient}

In Section \ref{sec:metallicity} we demonstrated that the WISE W1 and
W2 bands provide a very good measure of an M-giant's
metallicity. Given this fact, we now investigate the metallicity
gradient in the stream. Previous attempts to do so have been
published, but are usually limited to small sample sizes or
are drawn from inhomogeneous populations
\citep[e.g.][]{Chou2007,Monaco2007,Chou2010,Keller2010}. Using our
photometric relation, we are able to estimate metallicities for a
large number of stars, all belonging to the same population
(i.e. M-giants).

\begin{figure}
\plotone{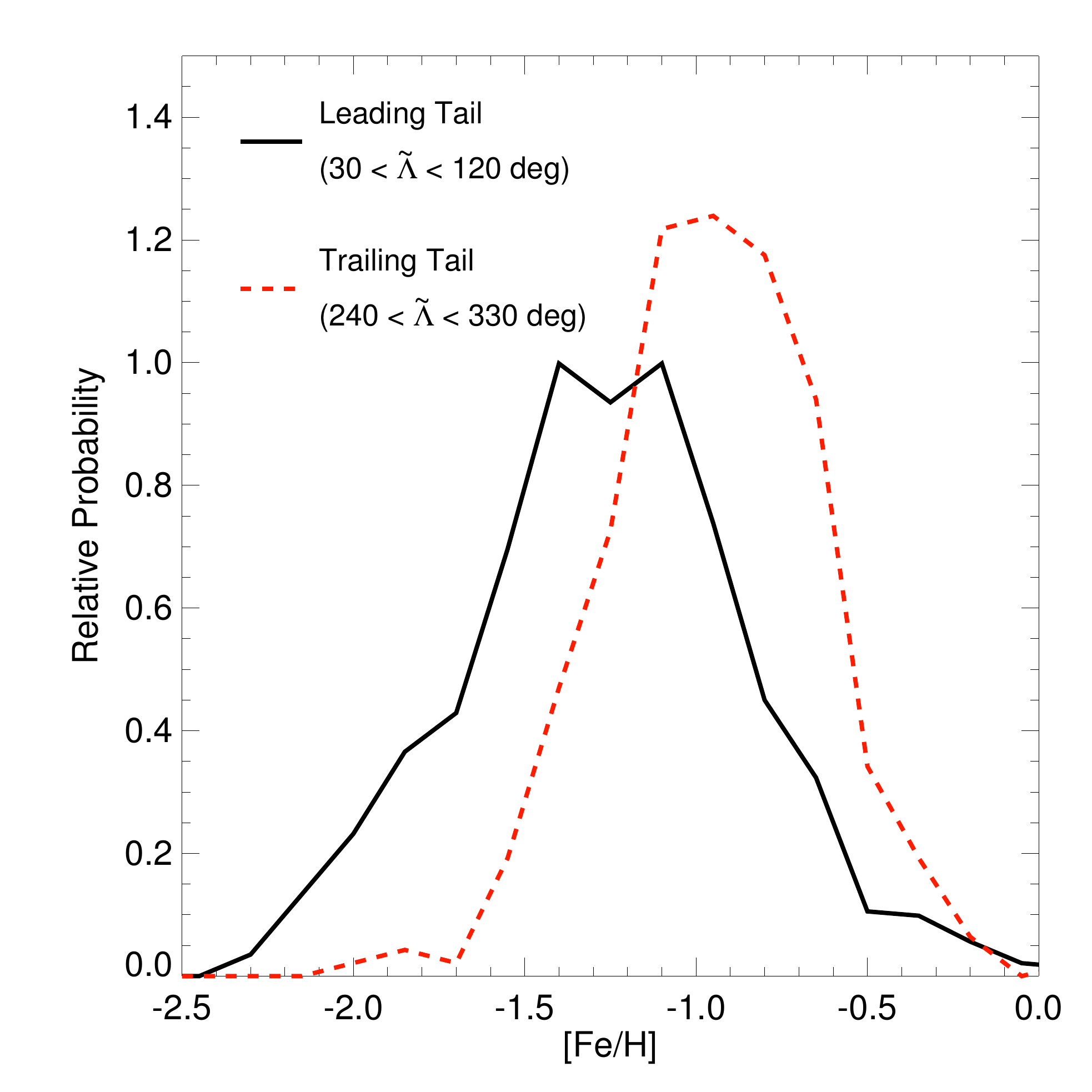}
\caption{Metallicity distributions for leading
($30<\tilde{\Lambda}<120$; solid) and trailing
($240<\tilde{\Lambda}<330$; dashed) tails of the Sagittarius stream,
estimated using the WISE M-giant metallicity relation defined in
Section \ref{sec:metallicity}. This plot only uses stars which are at
$-10<\tilde{B}<10$ and have distances within 10 kpc of the stream.}
\label{fig:sgr_metallicity}
\end{figure}

We have measured the metallicity distribution at two locations along
the stream, as shown in Figure \ref{fig:sgr_metallicity}. Here we have
restricted ourselves to only plotting stars with $-10<\tilde{B}<10$
deg and (as in the previous subsection) those whose distances are
within 10 kpc of the stream.
Although our photometric metallicity relation is
valid over a large range ($-1.5\la{\rm [Fe/H]}\la0$ dex), values
beyond this should be treated with caution; this is because our linear
relation is based on APOGEE data which only covers $-1.5\la{\rm
[Fe/H]}\la0$ dex, and so values beyond this are based on
extrapolations. As discussed in Section \ref{sec:metallicity}, the
uncertainty in this relation is 0.22 dex.

From Figure \ref{fig:sgr_metallicity} we can see a clear offset
between the two streams, with the leading stream being more
metal-poor. This is in good agreement with the result in \citet[][see
their Figure 7]{Monaco2007}, albeit with an order of magnitude more
stars. Our measurement of the trailing tail metallicity is also in
good agreement with that derived from main-sequence turn-off stars in
\citet{Carlin2012}. The mean and standard deviation of our
distributions are ($-1.18$, 0.43) and ($-0.88$, 0.31) dex for the
leading and trailing streams, respectively. Since the uncertainty in
the photometric metallicity relation is around 0.2 dex, the small
dispersions indicate that there is little intrinsic spread in the
M-giant metallicity for each tail, especially for the trailing
tail. Note that we were unable to find any significant gradients {\it
within} the leading or trailing streams, probably due to the lack of
precision in the metallicity measurements.

Such an offset between the two streams can be explained by internal
metallicity gradients in the progenitor dwarf. If the outer-regions
are more metal-poor, then the fact that these will be stripped earlier
than the inner regions sets up a metallicity gradient along the
stream. Although our leading and trailing metallicity distributions
are both measured around 100 deg from the core, they actually
correspond to material stripped at different times. By analyzing the
Sgr disruption history from the model of \citet{Law2010}, we find
that our trailing material is predicted to consist of debris stripped
during the most-recent pericentric passage, while our leading material
includes debris stripped from the last two pericentric passages (see
Fig. 17 of \citealt{Law2010}). This naturally leads to the observed
offset in metallicity (and also explains the narrower dispersion in
metallicity for the trailing tail), from which we can conclude that
the progenitor dwarf must have been massive enough to support internal
metallicity gradients.

One may wonder whether our distributions are affected by interlopers
from the aforementioned Tri-And system. However, if we minimize
potential Tri-And contamination by avoiding $20<RA<60$ deg
\citep{Deason2014}, then our findings remain unchanged, indicating
that this structure is not strongly biasing our results. 
Contamination from foreground M-dwarfs could also affect our
results, but if we reduce potential contamination by employing a
stricter $(J-K)_{0}$ cut of 0.95 mag (see Figure \ref{KM}) our
findings are unchanged. 

\section{Conclusions}
\label{sec:conclusions}

We have shown that WISE+2MASS photometry can be used to efficiently
identify M-giant stars. By using a large sample of spectroscopically
confirmed M-giants from LAMOST we have devised a photometric selection
(Equation \ref{eq:selection}) that contains only a few per cent dwarf
contamination and negligible contribution from QSOs. At fainter
magnitudes the dwarf contamination will increase, but as we have
shown in Section \ref{sec:photometry} this method is far superior to
those constructed using 2MASS photometry alone.

We have also constructed new photometric distance relations, using
large samples of M-giants from known structures
(e.g. LMC/SMC/Sgr). The variations in these distance relations
reflects the differing chemical composition of these
structures. Unlike most previous attempts to define distance
relations, our approach is entirely empirical and does not rely on
stellar models. We have demonstrated the accuracy of these relations
by investigating the tilt of the LMC. The fact that we are able to
reproduce the previously detected tilt shows that when averaging over
large samples of M-giants, these relations can be used to obtain precise
distances.

As has been noted previously
\citep[e.g.][]{Koposov2014,Schlaufman2014}, the WISE IR bands are
sensitive to a star's metallicity. Using the high-S/N,
high-resolution spectroscopic metallicities from APOGEE, we found a
strong correlation between the $({\it W1}-{\it W2})_{0}$ color and 
$\rm [M/H]$. This is well-fit by a linear relation (Equation
\ref{eq:metallicity}), with a scatter of 0.35 dex. The most remarkable
aspect of this is that the scatter in $({\it W1}-{\it W2})_{0}$ is
comparable to the observational error in the photometry, implying that
the intrinsic dispersion in this relation must be exceptionally
small. This means that it is possible to quickly amass large samples
of M-giants which are (a) free from dwarf contamination and (b) have
good metallicity estimates, without having to undertake time consuming
spectroscopic observations.

We applied our photometric selection criteria to the ALLWISE catalog
to obtain a large sample of stars from the Sgr stream. Using the above
distance and metallicity relations, we then investigated the
properties of the stream. We confirmed that the plane of the orbit is
mis-aligned for the leading and trailing tails, as is the case for the
main-sequence turn-off population. The recent detection of Sgr debris
towards the Galactic anti-center \citep{Koposov2014} was confirmed and,
although the WISE data does not allow us to detect the recently
identified apocenter of the trailing tail \citep{Belokurov2014}, the
anti-center material clearly forms a bridge between the known trailing
tail and the proposed apocenter detection. The proposed Branch C of
the stream can also be seen in our M-giant sample and, despite being
relatively tentative, the fact that it has been seen in a range of
tracers (subgiants, red clump stars and now M-giants) means that it's
unlikely to be spurious. However, we are unable to confirm that Branch
C is an extension of the trailing tail as we find no stars connecting
this to the trailing apocenter material.

Our photometric metallicity relation also allowed us to investigate
the metallicity of the stream. Using a large sample of stars from a
homogeneous population alleviates biases inherent when using a variety
of different tracer populations. The offset between the leading and
trailing material can be clearly seen (Figure
\ref{fig:sgr_metallicity}) and is around 0.2 dex. The fact that there
is an offset is consistent with prediction from the model of
\citet{Law2010}, as our leading material sample consists of debris
stripped in the last two pericentric passages, while our trailing
sample consists only of debris stripped in the most recent pericentric
passage. There can now be no doubt that the progenitor system must
have been massive, containing a significant radial metallicity
gradient, otherwise such metallicity offsets would not be present
along the stream.

M-giants are a valuable resource for studying our Galaxy. Their bright
absolute magnitudes means that they can be used to trace distant
structures in the disc and halo. In future work we will investigate the
disc population, studying low-latitude substructures or the
truncation of the stellar disc. A more complete picture can be inferred
when radial velocities are incorporated into the analysis. For
example, the first data release of the LAMOST survey contains  almost
10,000 M-giant spectra, which we are now in the process of
analyzing. This will be especially important for dissecting
low-latitude substructures, but will also allow us to confirm the 
nature of Sgr Branch C and potentially detect new halo
overdensities. As well as substructures, M-giant velocities
can be used to study the global properties of the Milky Way, for
example constraining the mass profile of the halo.

\section*{Acknowledgments}

The authors wish to thank Sergey Koposov and Vasily Belokurov for
their input into this study. We are also indebted to James
R.A. Davenport for providing the extinction data from Figure 8 of
\cite{Davenport2014}. We wish to thank the referee, whose comments
helped improve the clarity of the work.

This work was supported by the following grants: the 973 Program 2014
CB845702/2014CB845700; the Strategic Priority Research Program ``The
Emergence of Cosmological Structures'' of the Chinese Academy of
Sciences (CAS, grant XDB09010100); the CAS One Hundred Talent fund;
and by the National Natural Science Foundation of China (NSFC) under
grants 11173002, 11333003, 11373054, 11073038, 11073001, and
11373010 and 11390373. JLC and HJN acknowledges support from the
U.S. National Science Foundation under grants AST 09-37523 and AST 14-09421.

ZJ would like to acknowledge the National Natural Science Foundation
of China (NSFC, Grant No. 11503066） and the Shanghai Natural Science
Foundation (14ZR1446900).

The Guoshoujing Telescope (LAMOST) is a
National Major Scientific Project built by the CAS. Funding for the
project has been provided by the National Development and Reform
Commission. LAMOST is operated and managed by the National
Astronomical Observatories, CAS.

This publication makes use of data products from the Wide-field
Infrared Survey Explorer, which is a joint project of the University
of California, Los Angeles, and the Jet Propulsion
Laboratory/California Institute of Technology, funded by the National
Aeronautics and Space Administration.

\appendix

\section{Extinction correction}
\label{sec:extinction}

Since many of our M-giants are at low latitudes, we need to carefully
take extinction into account. Recent work by \citet{Davenport2014}
has shown that at near- and mid-infrared bands the dust extinction law
varies as a function of Galactic latitude. This can result in up to an
order of magnitude difference in $A_\lambda/A_r$ when comparing
high- to low-latitudes.

\begin{figure}
\plotone{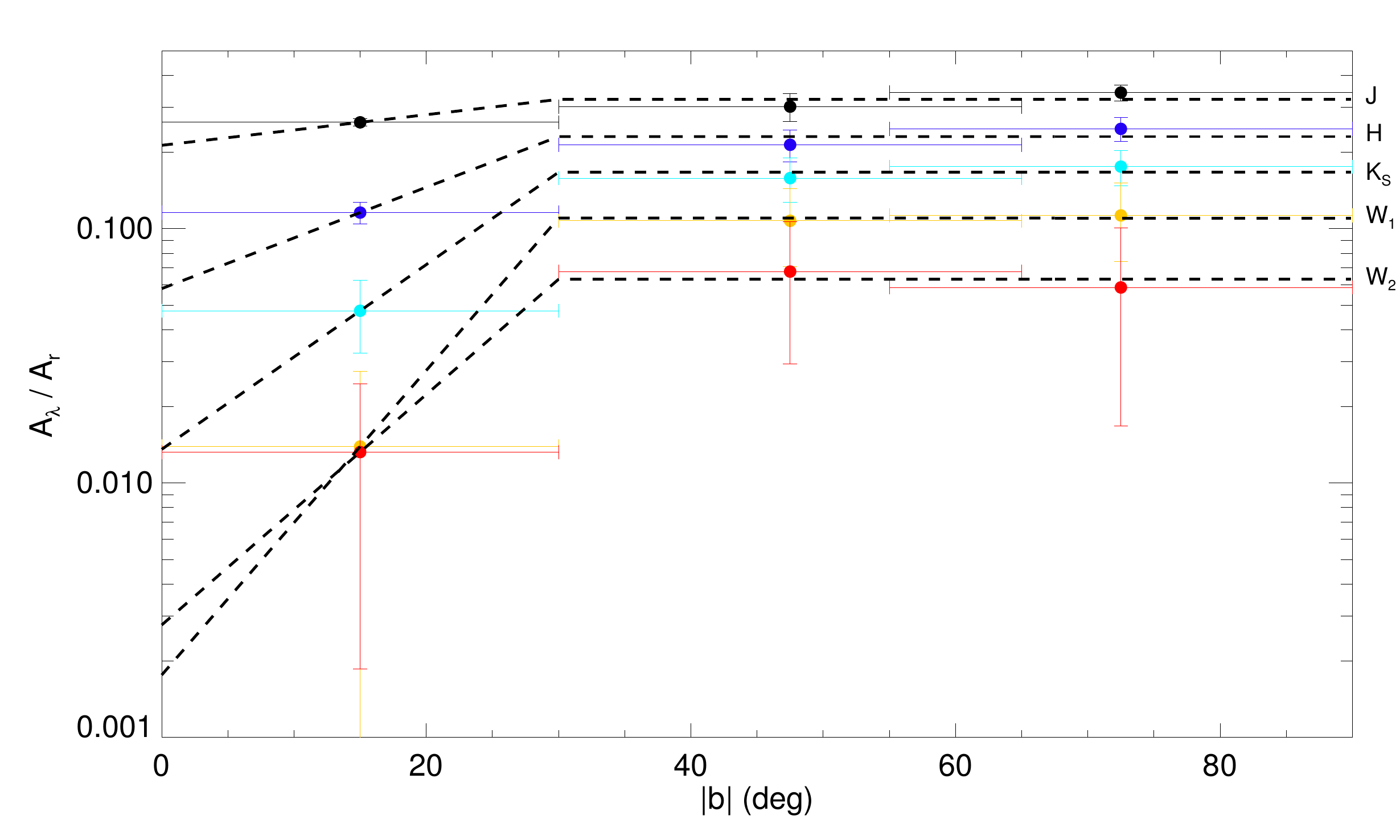}
\caption{Our latitude-dependent extinction law derived from the data
of \citet{Davenport2014}. The horizontal error bars denote the range
of $b$ used for that data-point.}
\label{fig:extinction}
\end{figure}

We account for this latitude dependence using a simple
prescription. Taking the data from Figure 8 of \citet{Davenport2014}, we
fit the extinction law for each band using a constant value for
$|b|>30$ deg and a power-law for $|b|<30$,
\begin{eqnarray}
\nonumber
A_\lambda/A_r \:\: &=& \:\: c \: 10^{m\left(\left|b\right|-30\right)}
\:\:\:\:\:{\rm for} \:\:\: \left|b\right| < 30 \: {\rm deg}\\
A_\lambda/A_r \:\: &=& \:\: c
\hspace{2.1cm}
{\rm for} \:\:\: \left|b\right| \ge 30 \: {\rm deg}.
\label{eq:extinction}
\end{eqnarray}
The resulting fits are shown in Figure \ref{fig:extinction} and the
values of $m$ and $c$ are given in Table \ref{tab:extinction}.
Throughout the paper we correct for extinction using Equation
(\ref{eq:extinction}).

\begin{table}
\begin{center}
\caption{The values of the parameters in our latitude-dependent
extinction law (Equation \ref{eq:extinction}), derived from the data
of \citet{Davenport2014}.}
\label{tab:extinction}
\begin{tabular}{lccc}
\hline
$\lambda$ & $m$ & $c$ \\
\hline
$J$ & 0.0060&    0.3223\\
$H$ & 0.0199&    0.2302\\
$K$ & 0.0363&    0.1666\\
$W1$ & 0.0599&   0.1100\\
$W2$ & 0.0453&   0.0633\\
\hline
\end{tabular}
\end{center}
\end{table}

\section{Table of halo M-giants}
\label{sec:online_table}

Here we present a table of halo M-giants. These have been
selected using the photometric quality cuts described in Section
\ref{sec:WISE} and the M-giant classification from Equation
(\ref{eq:selection}). We impose a further cut $J_0 > 12$ mag, to
ensure that our stars are at distances of greater than $\sim$ 10 kpc
and hence remove nearby disc stars.
Distances for these stars can be calculated using Equation
\ref{eq:distance}, metallicities using Equation \ref{eq:metallicity},
and extinctions using Equation \ref{eq:extinction}. Values of $E(B-V)$
are derived from the maps of \citet{Schlegel1998}.
This table must be used with caution at low-latitudes, since source
confusion and problems with the extinction maps (e.g. the low
resolution) can affect the reliability of the sample.

\begin{table*}
\begin{center}
\caption{Table of positions, magnitudes and reddening for a sample of
  halo M-giants. The full table is available in the online version;
  here we present only the first two rows.}
\label{tab:online}
\begin{tabular}{lccccccccccc}
\hline
ALLWISE ID & RA & Dec & $J$ & $\delta J$ & $K$ & $\delta K$ & $W1$ & $\delta W1$ & $W2$ &
$\delta W2$ & $E(B-V)$\\
& (deg) & (deg) & mag & mag & mag & mag & mag & mag & mag & mag & mag\\
\hline
J010035.43-021511.8 & 15.147661 & -2.253298 &  14.583  & 0.025 &  13.441  & 0.022 &  13.034  & 0.022 &  13.057  & 0.019 &  0.0367\\
J024752.75-003735.2 & 41.969832 & -0.626461 &  13.892  & 0.023 &  12.991  & 0.023 &  12.900  & 0.020 &  13.009  & 0.018 &  0.0356\\
... &... &... &... &... &... &... &... &... &... &...\\
\hline
\end{tabular}
\end{center}
\end{table*}

\end{document}